\documentclass[prb,floats,aps,amssymb,preprint,epsf,epsfig,rotate]{revtex4}

\begin{document}
\title{Exact-exchange Kohn-Sham potential, surface energy, and work function of
jellium slabs}

\author{C. M. Horowitz$^{1,2}$, C. R. Proetto$^{3,4,*}$, and J. M. Pitarke$^{5,6}$}

\affiliation{$^1$Donostia International Physics Center (DIPC), E-20018 San Sebastian, Spain.}
\affiliation{$^2$Instituto de Investigaciones Fisicoqu\'imicas Te\'oricas 
y Aplicadas, (INIFTA), UNLP, CCT La Plata-CONICET. C. C. 16 Suc. 4, 1900,
La Plata. Argentina.}
\affiliation{$^3$Instit\"ut f\"ur Teoretische Physik, Freie Universitat Berlin, Arnimallee 14, 
D-14195 Berlin, Germany.}
\affiliation{$^4$European Theoretical Spectroscopy Facility (ETSF).}
\affiliation{$^5$CIC nanoGUNE Consolider, Mikeletegi Pasealekua 56, E-2009 Donostia, Basque Country}
\affiliation{$^6$Materia Kondentsatuaren Fisika Saila, UPV/EHU, and Centro F\'isica Materiales CSIC-UPV/EHU, 
644 Posta kutxatila, E-48080 Bilbo, Basque Country}

\date\today

\begin{abstract}
Exact-exchange self-consistent calculations of the Kohn-Sham potential, surface
energy, and work function of jellium slabs are reported in the framework of
the Optimized Effective Potential (OEP) scheme of Density Functional Theory.
In the vacuum side of the jellium surface and at a distance $z$ that is
larger than the slab thickness, the exchange-only Kohn-Sham potential is found
to be image-like ($\sim -e^2/z$) but with a coefficient that differs from that
of the classical image potential $V_{im}(z)=-e^2/4z$. The three OEP contributions
to the surface energy (kinetic, electrostatic, and exchange) are found to
oscillate as a function of the slab thickness, as occurs in the case of the
corresponding calculations based on the use of single-particle orbitals and
energies obtained in the Local Density Approximation (LDA). The OEP work
function presents large quantum size effects that are absent in the LDA and
which reflect the intrinsic derivative discontinuity of the exact Kohn-Sham
potential.
\end{abstract}


\maketitle

\section{Introduction}

The analysis of the electronic structure of metal surfaces poses a big
theoretical challenge: a suitable calculational tool is needed for large,
interacting, and strongly inhomogeneous many-electron systems. More than
thirty years since its first application by Lang and Kohn to the surface
problem,\cite{lang,reviewlang} little doubt exists that one method of choice
for the fulfilling of this goal is Density Functional Theory (DFT).\cite
{parr,dreizler} DFT aims to a microscopic understanding of atoms, molecules,
clusters, surfaces, and bulk solids starting from the fundamental laws of
quantum mechanics. In the Kohn-Sham (KS) implementation of DFT,\cite{ks} the
complicated many-body problem is mapped to an effective single-particle
problem, with particles subjected to an effective single-particle potential
(the KS potential). Although this mapping is exact, it gives no clue as to
how to calculate in practice the so-called exchange-correlation ($xc$)
contribution to the KS potential. Lang and Kohn solved this problem by using
the Local Density Approximation (LDA) for the surface problem.
\cite{lang,reviewlang} In LDA, the $xc$ potential at each point is taken to be
that of a homogeneous interacting electron gas with the local density. Since
then, many authors have calculated the electronic properties of metal surfaces
by using either the LDA\cite{nekovee} or further elaborations that incorporate
non-local ingredients to the unknown $xc$ functional.\cite{zhang,constantin}
Other schemes of the computational electronic-structure tool kit available for
the investigation of solid surfaces are the Fermi hypernetted chain (FHNC)
method,\cite{kkq,kk} the $GW$ approximation,\cite{eguiluz} Quantum Monte Carlo
(QMC),\cite{li,acioli,wood} and the inhomogeneous Singwi-Tosi-Land-Sj\"olander
(ISTLS) approach.\cite{istls} 

In the framework of the Optimized Effective Potential (OEP) scheme of DFT,
\cite{grabo,engel} which had been first used in the context of atomic physics,
\cite{talman} correlation is ignored altogether and the exact-exchange KS
potential is obtained. Several advantages are associated with the use of the
exact-exchange energy functional of DFT: (i) it corrects the
self-interaction problem inherent in approximate treatments of the exchange
energy\cite{perdew} (this problem is particularly acute for localized
systems such as atoms and molecules, although it is not relevant for extended
systems like bulk solids and solid surfaces); (ii) it yields great improvements
in the study of the KS eigenvalue spectrum,\cite{dellasala} semiconductor band
structures and excitations,\cite{bylander} and nonlinear optical properties;
\cite{vangisbergen} (iii) it yields the correct asymptotics;\cite{kriebich}
(iv) it reproduces the derivative discontinuity which should be present in the
KS exchange potential each time the number of particles crosses through an
integer value;\cite{kli,pplb,pl,p,tozer} and (v) it yields the correct
two-dimensional (2D) exchange energy per particle in the case of a quasi-2D
electron gas.\cite{kim} It is the aim of this paper to provide {\it benchmark}
exact-exchange OEP calculations for jellium slabs, with the expectation that
more accurate DFT schemes that include correlation be developed by starting
from a well founded exchange analysis and tested once reduced to their
exchange-only ($x$-only) counterparts.

The rest of the paper is organized as follows: We give in Section II the
general theoretical background which will be used in the following sections;
Section III is devoted to a discussion of the asymptotic behaviour of the
exact-exchange KS potential of jellium slabs; in Sections IV and V
we give the results that we have obtained for the OEP surface energy and work
function, respectively, and in Section VI we present the conclusions.

\section{The OEP approach}

Our calculations are restricted to a jellium-slab model of metal surfaces,
where the discrete character of the positive ions inside the metal
is replaced by a uniform distribution of positive charge (the jellium).
The positive jellium density is defined as 
\begin{equation}
n_{+}(z)=\overline{n}\,\theta \left( \frac{d}{2}-\left| z+\frac{d}{2}\right|
\right),
\label{jellium}
\end{equation}
which describes a slab of width $d$, number density $\overline{n}$,\cite{note0}
and jellium edges at $z=-d$ and $z=0$; $\theta(x)$ represents the Heaviside
step function: $\theta (x)=1$ if $x > 0$ and $\theta (x)=0$ if $x < 0$. A
schematic view of our jellium slab is given in Fig.~1. Besides, and for
convenience for the numerical calculations, infinite barriers are located far
from the jellium edges, well inside the left and right evanescent vacuum
regions. We have checked that these infinite barriers are located far enough
for all the numerical calculations presented here to be independent of their
precise location. \cite{note0a} The jellium-slab model is invariant under translations in
the $x-y$ plane, so the KS eigenfunctions can be factorized as follows 
\begin{equation}
\varphi_{i,{\bf k}}({\bf r})=\frac{e^{i{\bf k\cdot \rho }}}{\sqrt{A}}\,
\xi_{i}(z),
\label{KSfunctions}
\end{equation}
where ${\bf \rho }$ and ${\bf k}$ are the in-plane coordinate and
wave-vector, respectively, and $A$ represents a normalization area.
$\xi _{i}(z)$ are the normalized spin-degenerate
eigenfunctions for electrons in slab discrete levels (SDL) $i$ $(i=1,2,...)$
with energy $\varepsilon_{i}.$ They are the solutions of the effective one-dimensional
KS equation 

\begin{equation}
\widehat{h}_{\text{KS}}^{i}(z)\xi _{i}(z)=\left[ -\frac{\hbar ^{2}}{2m_{e}}
\frac{\partial ^{2}}{\partial z^{2}}+V_{\text{KS}}\left( z\right)
-\varepsilon _{i}\right] \xi _{i}(z)=0,
\label{KSequations}
\end{equation}

with $m_e$ the bare electron mass.

The KS potential $V_{\text{KS}}$ entering Eq.~(\ref{KSequations}) is the sum of
two distinct contributions: 
\begin{equation}
V_{\text{KS}}(z)=V_{\text{H}}(z)+V_{xc}(z),  \label{KSpotential}
\end{equation}
where $V_{\text{H}}(z)$ is the classical (electrostatic) Hartree potential,
given by \cite{hartree.a} 
\begin{equation}
V_{\text{H}}(z)=-2\pi e^{2}\int_{-\infty }^{\infty }dz^{\prime }\left|
z-z^{\prime }\right| \left[ n(z^{\prime })-n_{+}(z^{\prime })\right] .
\label{hartree}
\end{equation}

Here, $n(z)$ is the electron number density \cite{dens.aa} 
\begin{equation}
n(z)=\frac{1}{2\pi }\sum_{i}^{occ.}\left( k_{F}^{i}\right) ^{2}\left| \xi
_{i}(z)\right| ^{2},  \label{density}
\end{equation}
where
$k_{F}^{i}=\sqrt{2m_{e}(\mu -\varepsilon _{i})}/\hbar$,
and $\mu=\mu (\bar n,d) $ is the chemical potential, 
which in turn is determined from the neutrality
condition for the whole system by the condition
$\sum_{i}^{occ.}(k_{F}^{i})^{2}=2\pi \,d\,\overline{n}$. $V_{xc}(z)$ is the
nonclassical $xc$ potential, which is obtained as the functional derivative of
the so-called $xc$ energy functional $E_{xc}[n(z)]$:\cite{factor.a}
\begin{equation}
V_{xc}(z)\equiv \frac{1}{A}\frac{\delta E_{xc}[n(z)]}{\delta n(z)}.
\label{xcpotential}
\end{equation}

Applications of DFT typically proceed from {\it explicit} density-dependent
forms of $E_{xc}$, as obtained using a variety of local or semi-local
approximations. However, in the last few years increasing attention has been
devoted to orbital-dependent forms of $E_{xc}$:
$E_{xc}=E_{xc}\left[ \left\{\xi_{i}\right\},\left\{\varepsilon_{i}\right\}
\right]$, which are only {\it implicit} functionals of the electron density
$n(z)$. In this case, one resorts to the OEP method\cite{grabo} or,
equivalently, uses repeatedly the chain rule for functional derivatives to
obtain the following expression for the $xc$ potential of
Eq.~(\ref{xcpotential}):\cite{exchangefunctional}
\begin{equation}
V_{xc}(z)=\frac{1}{A}%
\sum_{i}^{occ.}\int\limits_{-\infty }^{\infty }dz^{\prime
}\int\limits_{-\infty }^{\infty }dz^{^{\prime \prime }}\left[ \frac{\delta
E_{xc}}{\delta \xi _{i}(z^{^{\prime \prime }})}\frac{\delta \xi
_{i}(z^{^{\prime \prime }})}{\delta V_{\text{KS}}(z^{\prime })}+\text{c.c.}%
\right] \frac{\delta V_{\text{KS}}(z^{\prime })}{\delta n(z)}.
\label{chainrule}
\end{equation}
Multiplying Eq.~(\ref{chainrule}) by the KS density-response function
$\chi_{\text{KS}}^{}(z,z^{\prime })\equiv \delta n(z)/\delta V_{\text{KS}}(z^{\prime })$,
using the identity
\begin{equation}
\int\limits_{-\infty }^{\infty }\chi_{\text{KS}}^{} (z,z^{\prime })\;\chi_{\text{KS}} ^{-1}
(z^{\prime},z^{^{\prime \prime }})\;dz=\delta (z-z^{^{\prime \prime }}), 
\label{delta}
\end{equation}
comparing Eqs.~(\ref{xcpotential}) and (\ref{chainrule}), and integrating over
the coordinate $z$, one finds
\begin{equation}
\int\limits_{-\infty }^{\infty }\frac{\delta E_{xc}}{\delta n(z^{\prime })}%
\;\chi_{\text{KS}}^{}(z,z^{\prime })\;dz^{\prime }=\sum_{i}^{occ.}\int\limits_{-\infty
}^{\infty }\left[ \frac{\delta E_{xc}}{\delta \xi _{i}(z^{\prime })}\frac{%
\delta \xi _{i}(z^{\prime })}{\delta V_{\text{KS}}(z)}+\text{c.c.}\right]
\;dz^{\prime }.
\label{oep}
\end{equation}

The nice feature of Eq.(\ref{oep}) is that
$\delta \xi_{i}(z^{\prime })/\delta V_{\text{KS}}(z)$ and
$\chi_{\text{KS}}^{}(z,z^{\prime})$ are simply obtained from the solutions of
Eq.~(\ref{KSequations}), as follows 
\begin{equation}
\frac{\delta \xi _{i}(z^{\prime })}{\delta V_{\text{KS}}(z)}=\xi
_{i}(z)\sum_{j\,(\neq i)}\frac{\xi _{j}(z^{\prime })^{*}\xi _{j}(z)}{\left(
\varepsilon _{i}-\varepsilon _{j}\right) }\equiv \xi
_{i}(z)\;G_{i}^{\text{KS}}(z^{\prime },z),
\label{pt1}
\end{equation}
and 
\begin{eqnarray}
\chi_{\text{KS}}^{}(z,z^{\prime }) &=&\sum_{i}^{occ.}\int\limits_{-\infty }^{\infty
}dz^{^{\prime \prime }}\left[ \frac{\delta n(z)}{\delta \xi _{i}(z^{^{\prime
\prime }})}\frac{\delta \xi _{i}(z^{^{\prime \prime }})}{\delta
V_{\text{KS}}(z^{\prime })}+\text{c.c.}\right] ,  \label{responsefunction1} \\
&=&\frac{1}{4\pi }\sum_{i}^{occ.}\left[ \left( k_{F}^{i}\right) ^{2}\xi
_{i}(z)^{*} \xi _{i}(z^{\prime })G_{i}^{\text{KS}}(z^{\prime },z)+\text{c.c.}\right],
\label{responsefunction2}
\end{eqnarray}
where $G_{i}^{\text{KS}}(z^{\prime },z)$ is the Green function of noninteracting KS
electrons. In the calculation of $\chi_{\text{KS}}^{}(z,z^{\prime })$, the chain rule
for functional
derivatives has been used; now we are considering the density itself as a functional
of the occupied SDL. In obtaining Eq.~(\ref{responsefunction2}) from
Eq.~(\ref{responsefunction1}), we have used Eq.~(\ref{pt1}) and also that 
\[
\frac{\delta n(z)}{\delta \xi _{i}(z^{\prime })}=\delta (z-z^{\prime })\frac{%
\left( k_{F}^{i}\right) ^{2}}{2\pi }\xi _{i}(z)^{*}, 
\]
which follows from Eq.~(\ref{density}).

Introducing Eqs.~(\ref{pt1}) and (\ref{responsefunction2}) into the
central Eq.~(\ref{oep}), we obtain the final and
compact version of the OEP integral equation for $V_{xc}(z)$:
\begin{equation}
\sum_{i}^{occ.}S_{i}(z)=0,
\label{oepcompact}
\end{equation}
where 
\[
S_{i}(z)=\left( k_{F}^{i}\right) ^{2}\Psi _{i}(z)^{*}\xi _{i}(z)+\text{c.c.},
\]
and 
\begin{equation}
\Psi _{i}(z)=\sum_{j\,(\neq i)}\frac{\xi _{j}(z)}{\left( \varepsilon
_{i}-\varepsilon _{j}\right) }\int\limits_{-\infty }^{\infty }\xi
_{j}(z^{\prime })^{*}\,\Delta V_{xc}^{i}(z^{\prime })\,\xi _{i}(z^{\prime
})dz^{\prime }.
\label{shifts}
\end{equation}
Here, $\Delta V_{xc}^{i}(z)=V_{xc}(z)-u_{xc}^{i}(z)$, where $u_{xc}^{i}(z)$ are
SDL-dependent $xc$ potentials of the form:\cite{ux.explicito}
\[
u_{xc}^{i}(z)\equiv \left[ 4\pi /A(k_{F}^{i})^{2}\xi _{i}(z)^{*}\right]
\delta E_{xc}/\delta \xi _{i}(z).
\]

The magnitudes $\Psi _{i}(z)$ are called the ``shifts'', as they can be
physically interpreted as the first-order corrections of the KS eigenfunctions
$\xi_{i}(z)$ under the perturbation $\Delta V_{xc}^{i}(z)$. These
shifts also provide a useful and practical tool for the numerical solution
of the OEP equation.\cite{kummel,rigamonti1} From Eq.~(\ref{shifts}), we find
the orthogonality constraint between the KS eigenfunctions and the shifts:
$\int \xi _{i}(z)^{*}\,\Psi _{i}(z)\,dz=0$. It is also
immediate that the shifts are invariant under the replacement $%
V_{xc}(z)\rightarrow V_{xc}(z)+\alpha,$ with $\alpha$ being an arbitrary
constant. This means that the above set of equations determines $V_{xc}(z)$
up to an additive constant, which should be fixed by imposing a suitable
boundary condition. Moreover, the shifts $\Psi _{i}(z)$ are easily found to
satisfy the following inhomogeneous differential equation:\cite{dellasala} 

\begin{equation}
\widehat{h}_{\text{KS}}^{i}(z)\Psi _{i}(z)=-\left[ \Delta
V_{xc}^{i}(z)-\Delta \overline{V}_{xc}^{i}\right] \xi _{i}(z).
\label{de-shifts}
\end{equation}
Here, mean values are defined as
$\overline{O}^{i}=\int \xi _{i}(z)^{*}O^{i}(z)\,\xi _{i}(z)\,dz.$

Equations~(\ref{KSequations})-(\ref{hartree}) and (\ref{oepcompact}), which
determine the local $V_{xc}(z)$ corresponding to a given SDL-dependent
$E_{xc}$, form a closed system of equations (the OEP equations), which should
be solved in a self-consistent way. In order to accomplish some contact with
other useful versions of the OEP equations for the present problem, a few
additional steps are required.
First of all, we write 
\[
\xi _{i}(z)^{*}\widehat{\,h}_{\text{KS}}^{i}(z)\,\Psi _{i}(z)=-\frac{\hbar
^{2}}{2m_{e}}\left[ \xi _{i}(z)^{*}\,\frac{\partial ^{2}\Psi _{i}(z)}{%
\partial z^{2}}-\Psi _{i}(z)\frac{\partial ^{2}\xi _{i}(z)^{*}}{\partial
z^{2}}\right] , 
\]
which is easily obtained from Eq.~(\ref{KSequations}). Secondly, we multiply
the left hand-side of Eq.~(\ref{de-shifts}) by $\xi _{i}(z)^{*}$ to obtain 
\begin{equation}
\frac{\hbar ^{2}}{2m_{e}}\left[ \xi _{i}(z)^{*}\,\frac{\partial ^{2}\Psi
_{i}(z)}{\partial z^{2}}-\Psi _{i}(z)\frac{\partial ^{2}\xi _{i}(z)^{*}}{%
\partial z^{2}}\right] =\left[ \Delta V_{xc}^{i}(z)-\Delta \overline{V}%
_{xc}^{i}\right] \left| \xi _{i}(z)\right| ^{2}.  \label{equality}
\end{equation}
Then, we start from the self-evident identity 
\begin{equation}
V_{xc}(z)=\sum_{i}^{occ.}\frac{\left( k_{F}^{i}\right) ^{2}\left| \xi
_{i}(z)\right| ^{2}}{4\pi n(z)}\left[ u_{xc}^{i}(z)+\Delta \overline{V}%
_{xc}^{i}+\Delta V_{xc}^{i}(z)-\Delta \overline{V}_{xc}^{i}+\text{c.c.}%
\right],
\label{identity}
\end{equation}
we eliminate the factor
$\left[ \Delta V_{xc}^{i}(z)-\Delta \overline{V}_{xc}^{i}\right]
\left| \xi _{i}(z)\right| ^{2}$ by using Eq.~(\ref{equality}), and we obtain 
\begin{equation}
V_{xc}(z)=\sum_{i}^{occ.}\frac{\left( k_{F}^{i}\right) ^{2}}{4\pi n(z)}%
\left\{ \left| \xi _{i}(z)\right| ^{2}\left[ u_{xc}^{i}(z)+\Delta \overline{V%
}_{xc}^{i}\right] +\frac{\hbar ^{2}}{2m_{e}}\left[ \xi _{i}(z)^{*}\,\frac{%
\partial ^{2}\Psi _{i}(z)}{\partial z^{2}}-\Psi _{i}(z)\frac{\partial
^{2}\xi _{i}(z)^{*}}{\partial z^{2}}\right] +\text{c.c.}\right\}.
\label{oeppotential}
\end{equation}
Finally, we proceed with the elimination from Eq.~(\ref{oeppotential}) of the
term proportional to $\partial ^{2}\Psi _{i}(z)/\partial z^{2}$, the subsequent
elimination of $\partial^{2}\xi _{i}(z)^{*}/\partial z^{2}$ proceeds via the
KS equations, and  as a result of all these manipulations we obtain the
following expression for the DFT $xc$ potential:
\cite{horowitz} 
\begin{equation}
V_{xc}(z)=V_{xc,1}(z)+V_{xc,2}(z),
\label{oeppotentialdecomposition}
\end{equation}
where 
\[
V_{xc,1}(z)=\sum_{i}^{occ.}\frac{\left( k_{F}^{i}\right) ^{2}\left| \xi
_{i}(z)\right| ^{2}}{4\pi n(z)}\left\{ u_{xc}^{i}(z)+\Delta \overline{V}%
_{xc}^{i}+\text{c.c.}\right\} 
\]
and 
\[
V_{xc,2}(z)=-\frac{1}{2\pi n(z)}\sum_{i}^{occ.}\left( \mu -\varepsilon _{i}\right) \left[ \left( k_{F}^{i}\right) ^{2}\Psi
_{i}(z)\xi _{i}(z)^{*}+\Psi _{i}^{\prime }(z)\xi _{i}^{\prime }(z)^{*}+\text{%
c.c.}\right],
\]
with primes denoting derivatives with respect to the $z$ coordinate. It is
important to note that Eqs.~(\ref{oepcompact}) and
(\ref{oeppotentialdecomposition}) are just two different, but fully
equivalent, ways to obtain the OEP $xc$ potential for
the present problem. If the shifts $\Psi_i(z)$ are (arbitrarily) forced to be
identically equal to zero, the only term that survives is $V_{xc,1}(z)$. This
is exactly the KLI approximation,\cite{kli} which brings the identification  
 $V_{xc,1}(z) \equiv V_{xc}^{KLI}(z)$. \cite{kli.2} As before,
Eqs.~(\ref{KSequations})-(\ref{hartree}) and
(\ref{oeppotentialdecomposition}) form a closed set
of equations, which should be solved self-consistently.

Both exchange and correlation have been included so far. Unless stated
otherwise, we will now focus on the $x$-only case,
where $E_{xc}$, $V_{xc}(z)$, and $u_{xc}(z)$ are replaced 
by $E_x$, $V_x(z)$, and $u_x(z)$, respectively. 
We have achieved the self-consistent
numerical solution of the $x$-only version of the OEP equations by two
different methods: $i)$ direct calculation of the shifts of
Eq.~(\ref{shifts}), by solving Eq.~(\ref{de-shifts}),\cite{kummel} and $ii)$
direct solution of the OEP integral equation for $V_{x}(z),$ as given by the
$x$-only version of Eq.~(\ref{oepcompact}).\cite
{rigamonti1} Both methods yield results that agree within numerical accuracy,
although the first approach is found to be computationally more efficient than
the second. Both methods face numerical instabilities beyond a critical
coordinate $z$ in the vacuum region. 

Finally, we note that the exact-exchange energy of a jellium slab is given by
the following expression:
\begin{equation}
E_x(d)={A\over 4\pi}\sum_i^{occ.} \left(k_F^i\right)^2 \int
\limits_{-\infty}^\infty dz\,|\xi_{i}(z)|^2 u_x^i(z),
\label{ex}
\end{equation}
where $u_x^i(z)$ represent the SDL-dependent exchange potentials
\begin{equation}
u_{x}^{i}(z)=-\frac{2e^{2}}{\left(k_{F}^{i}\right)^2}
\sum_{j}^{occ.}\frac{\xi _{j}(z)^{*}}{\xi _{i}(z)^{*}}\int\limits_{-\infty
}^\infty dz^\prime\, \frac{\xi _{i}(z^{\prime })^{*}\,
g(\Delta z\,k_{F}^{i},\Delta z\,k_{F}^{j})\,\xi _{j}(z^{\prime})}{(\Delta z)^3},
\label{odp}
\end{equation}
with $\Delta z=|z-z'|$, 
\begin{equation}
g(s,s')=s\,s'\,
\int\limits_{0}^{\infty }\frac{J_{1}(s\,t)J_{1}(s'\,t)}
{\sqrt{1+t^{2}}}\;\frac{dt}{t}
\label{odp2}
\end{equation}
being the ``universal'' (that is, independent of $V_{\text{KS}}$) function
introduced by Kohn and Mattsson,\cite{mattsson} and $J_{1}(x)$ being the
first-order cylindrical Bessel function.\cite{abra} 

\section{Asymptotics of the exact-exchange KS potential}

The long-range behavior of $V_{xc}(z)$ in the vacuum region is an important
and open issue in DFT studies of metal surfaces.\cite{jung} The aim of this 
section is to present a detailed derivation of the analytical asymptotic limit
of $V_{x}(z)$ reported in Ref.~\onlinecite{horowitz} for a slab geometry.
First of all, we note that by making the choice that
$V_{\text{KS}}(z\rightarrow \infty )\rightarrow 0,$ Eq.~(\ref{KSequations})
leads us to the conclusion that
$\xi _{i}(z\rightarrow \infty)\rightarrow e^{-\,z\,\sqrt{-2\,m_{e}\,\varepsilon
_{i}}\,/\,\hbar }$ for all occupied $i$ (disregarding a factor involving
powers of $z$). We also remark the following points: $i)$ Due to the
exponential decay of
$V_{\text{H}}(z\rightarrow \infty ),$ the assumption
$V_{\text{KS}}(z\rightarrow \infty )\rightarrow 0$ implies that
$V_{x}(z\rightarrow \infty )\rightarrow 0;$ $ii)$ for this choice of the zero
of energy, one finds $\varepsilon_{i}<0$ for all occupied states; $iii)$ the
slowest decaying of all the occupied SDL corresponds to $i=m,$ where $m$ is
the highest occupied SDL.

Now we look at the asymptotic behavior of the shifts $\Psi_i(z)$. Turning to
the $x$-only version of Eq.~(\ref{de-shifts}), 
\begin{equation}
\left[ -\frac{\hbar ^{2}}{2m_{e}}\frac{\partial ^{2}}{\partial z^{2}}+V_{%
\text{H}}\left( z\right) +V_{x}(z)-\varepsilon _{i}\right] \Psi
_{i}(z)=-V_{x}(z)\xi _{i}(z)+u_{x}^{i}(z)\xi _{i}(z)+\Delta \overline{V}%
_{x}^{i}\xi _{i}(z),  \label{de-xshifts}
\end{equation}
we focus on the asymptotic behavior of the three terms on the r.h.s. of this
equation:\cite{dellasala} 
\begin{mathletters}
\begin{eqnarray}
V_{x}(z &\rightarrow &\infty )\,\xi _{i}(z\rightarrow \infty )\rightarrow
V_{x}(z\rightarrow \infty )\,e^{-z\beta _{i}},  \label{first} \\
u_{x}^{i}(z &\rightarrow &\infty )\,\xi _{i}(z\rightarrow \infty
)\rightarrow e^{-z\beta _{m}},  \label{second} \\
\Delta \overline{V}_{x}^{i}\,\xi _{i}(z &\rightarrow &\infty )\rightarrow
e^{-z\beta _{i}},
\label{third}
\end{eqnarray}
with $\beta _{i}=\sqrt{-2m_{e}\varepsilon _{i}}/\hbar .$
Eq.~(\ref{second}) follows from an inspection of Eq.~(\ref{odp}) in the
limit $z\rightarrow \infty $: in this limit, the sum over $j$ is exponentially
dominated by the term $j=m$, and the result of Eq.~(\ref{second}) follows at
once. Hence, for $i\neq m$ Eq.~(\ref{de-xshifts}) yields
\end{mathletters}
\begin{equation}
\left[ -\frac{\hbar ^{2}}{2m_{e}}\frac{\partial ^{2}}{\partial z^{2}}%
-\varepsilon _{i}\right] \Psi _{i}(z\rightarrow \infty )\rightarrow
e^{-z\beta _{m}},
\end{equation}
i.e., $\Psi _{i}(z\rightarrow \infty )\rightarrow e^{-z\beta_{m}}.$
For $i=m$, all three terms in the r.h.s. of Eq.~(\ref{de-xshifts}) decay
equally (to exponential accuracy), and further analysis is necessary.
Eq.~(\ref{oepcompact}) can be rewritten as follows
\begin{equation}
\left( k_{F}^{m}\right) ^{2}\Psi _{m}(z)^{*}\xi _{m}(z)+\text{c.c.}%
=-\sum_{i=1}^{m-1}\left( k_{F}^{i}\right) ^{2}\Psi _{i}(z)^{*}\xi _{i}(z)-%
\text{c.c.,}
\end{equation}
and by studying its asymptotic limit, it is clear that its r.h.s. can be
approximated by the term $i=m-1$ (with exponential accuracy). Given that
both $\xi _{m}(z\rightarrow \infty )$ and $\Psi _{m-1}(z\rightarrow \infty )$
decay as $e^{-z\beta _{m-1}},$ it follows that $\Psi _{m}(z\rightarrow
\infty )$ decays as $\xi _{m-1}(z\rightarrow \infty )$, that is, $\Psi
_{m}(z\rightarrow \infty )\rightarrow e^{-z\beta _{m-1}}.$ Armed with these
results, the asymptotic limit of $V_{x}(z)$ is immediate from
Eq.~(\ref{oeppotentialdecomposition}): $V_{x,2}(z\rightarrow \infty )$ tends
exponentially to zero, while 
\begin{equation}
V_{x,1}(z\rightarrow \infty )\rightarrow u_{x}^{m}(z\rightarrow \infty
)+\Delta \overline{V}_{x}^{m}.  \label{kliasymp}
\end{equation}
The leading contribution to $u_{x}^{m}(z\rightarrow \infty )$ is easily
obtained from Eq.~(\ref{odp}), by considering once again that in this regime
the sum over $j$ is exponentially dominated by the term $j=m.$ For this
case, the integral over the coordinate $t$ can be evaluated analytically,
yielding 
\begin{equation}
u_{x}^{m}(z\rightarrow \infty )\rightarrow -e^{2}\int\limits_{-\infty
}^{\infty }\frac{\left| \xi _{m}(z^{\prime })\right| ^{2}}{\left|
z-z^{\prime }\right| }\,dz^{\prime }\left[ 1-\frac{I_{1}\left(
2k_{F}^{m}\left| z-z^{\prime }\right| \right) }{k_{F}^{m} \left| z-z^{\prime }\right| }%
+\frac{L_{1}\left( 2k_{F}^{m}\left| z-z^{\prime }\right| \right) }{ k_{F}^{m}\left|
z-z^{\prime }\right| }\right] ,  \label{sdpasymp1}
\end{equation}
where $I_{1}$ and $L_{1}$ are the modified Bessel and Struve functions,
respectively.\cite{abra} Noting now that in this regime $k_{F}^{m}\left|
z-z^{\prime }\right| \simeq k_{F}^{m}\,z\gg 1,$\cite{note} it is permissible
to expand the integrand of Eq.~(\ref{sdpasymp1}) as follows 
\begin{equation}
u_{x}^{m}(z\rightarrow \infty )\rightarrow -\frac{e^{2}}{z}%
\int\limits_{-\infty }^{\infty }\left| \xi _{m}(z^{\prime })\right|
^{2}\,dz^{\prime }\left[ 1+\frac{z^{\prime }}{z}-\frac{2}{\pi k_{F}^{m}z}+O(%
\frac{1}{z^{2}})\right] .  \label{sdpasymp2}
\end{equation}
Using the normalization of the orbitals $\xi_{m}(z)$, we obtain 
\begin{equation}
u_{x}^{m}(z\rightarrow \infty )\rightarrow -\frac{e^{2}}{z}\left( 1+\frac{%
\beta }{z}+...\right) ,  \label{sdpasymp}
\end{equation}
with $\beta =\overline{z}^m-2/\left( \pi k_{F}^{m}\right).$\cite{note.bb} Since the exchange
potential $V_x(z)$ has been chosen to vanish at large distances from the
surface into the vacuum [$V_x(\infty)=0$], Eq.~(\ref{kliasymp}) leads us to
the important constraint 
\begin{equation}
\Delta \overline{V}_{x}^{\,m}=\overline{V}_{x}^{\,m}-\overline{u}%
_{x}^{\,m}=0,  \label{bc}
\end{equation}
which fixes the undetermined constant in $V_{x}(z)$ discussed
above. All numerical results presented here have been obtained by using this
constraint. From Eqs.~(\ref{kliasymp}), (\ref{sdpasymp}), and (\ref{bc}), we
conclude that 
\begin{equation}
V_{x}(z\rightarrow \infty )\rightarrow V_{x,1}(z\rightarrow \infty
)\rightarrow u_{x}^{m}(z\rightarrow \infty )\rightarrow -\frac{e^{2}}{z}%
\left( 1+\frac{\beta }{z}+...\right) ,  \label{epasymp}
\end{equation}
which is the main result of this Section.

At this point, we emphasize that the asymptotics dictated by
Eq.~(\ref{epasymp}) hold only at $z$ coordinates that are larger than
$1/k_F^m$. As $k_F^m$ is of the order
of $1/d$ (or smaller, depending on the actual value of $d$),
Eq.~(\ref{epasymp}) shows that the $x$-only KS potential happens to be
four times larger than the classical image potential ($V_{im}(z)=-e^2/4z$) only
at a distance $z$ that is considerably larger than the slab thickness.
Furthermore, the arguments leading to
Eqs.~(\ref{second}) and (\ref{sdpasymp1}) are only valid for a discrete
slab spectrum, such that there is a finite energy gap between
$\varepsilon_{m}$ and the remaining occupied energy levels
$\varepsilon _{i}\;\left(i<m\right).$ An extension of the present OEP framework
to treat the case of a semi-infinite jellium surface\cite{nastos} is now in
progress\cite{new}.

Finally, we note that under the condition $V_{\text{KS}}(\infty)=0$
Eq.~(\ref{epasymp}) for the asymptotics of $V_{x}(z)$ remains valid when
correlation is included in the evaluation of the shifts $\Psi_i(z)$. The point
here is that the shifts are separable in their exchange and correlation
components, and they also satisfy separated differential equations (like
Eq.~(\ref{de-xshifts}) for exchange). Once exchange and correlation
contributions are splited, the analysis of the asymptotic behavior of
$V_x(z)$ follows the same lines as above, and the asymptotic limit of
Eq.~(\ref{epasymp}) remains the same.

\section{Surface energy}

In this section, surface-energy calculations are presented, as obtained at the
$x$-only level. The surface energy $\sigma $ is the work
required, per unit area of the new surface formed, to split the crystal in
two along a plane.\cite{lang} For our slab geometry, 
\begin{equation}
\sigma (d)=\frac{2E(d)-E(2d)}{2A},  \label{surfener}
\end{equation}
where $E(d)$ is the total ground-state energy for each half of the slab
after it is split (width $d$), and $E(2d)$ is the total ground-state energy
of the unsplit slab (width $2d$), both the split and unsplit systems
 with the same jellium density.

Following the standard DFT
energy-functional partitioning, 
the surface energy (without correlation contribution) can be written as the sum
of three terms,\cite{nekovee} 
\begin{equation}
\sigma (d)=\sigma _{K}(d)+\sigma _{el}(d)+\sigma _{x}(d),  \label{partition}
\end{equation}
where $\sigma _{K}(d)$ is the non-interacting kinetic contribution to the
surface energy, $\sigma _{el}(d)$ is the electrostatic surface energy due to
all non-compensated positive and negative charge distributions in the slab,
and $\sigma _{x}(d)$ is the exchange contribution to the surface energy.
From elementary physical arguments, it follows that $\sigma_K(d)<0,$
while $\sigma_{el}(d)$ and $\sigma_{x}(d)$ are both positive.\cite{reviewlang}
Also, the
stability of the slab against spontaneous fragmentation is accomplished if $%
\sigma (d)>0.$ From Eqs. (\ref{surfener}) and (\ref{partition}), one writes
\begin{equation}
 \sigma_{l}(d)=\left[ 2E_{l}(d)-E_{l}(2d)\right] /(2A),
\label{double.slab} 
\end{equation}
with $l=K,\;el,\,x,$ and
\begin{equation}
E_{K}(d)=\frac{A \hbar ^{2}}{4\pi m_{e}}\sum_{i}^{occ.}\left(
k_{F}^{i}\right) ^{2}\left[ \frac{\left( k_{F}^{i}\right) ^{2}}{2}%
-\int\limits_{-\infty }^{\infty }\xi _{i}(z)\frac{\partial ^{2}\xi _{i}(z)}{%
\partial z^{2}}\,dz\right] ,\label{kinetic}
\end{equation}
\begin{equation}
E_{el}(d)=\frac{A}{2}\int\limits_{-\infty }^{\infty }V_{\text{H}}(z)\left[
n(z)-n_{+}(z)\right] \,dz,  \label{electrostatic}
\end{equation}
and [see Eqs.~(\ref{ex})-(\ref{odp})] 
\begin{equation}
E_{x}(d)=-\frac{e^{2}A}{2\pi}\sum_{i,j}^{occ.}
\int\limits_{-\infty}^{\infty }dz
\int\limits_{-\infty}^{\infty }dz^\prime\,
\frac{\xi _{i}(z)^{*}\xi _{j}(z^{\prime })^{*}g(k_{F}^{i}\Delta
z,k_{F}^{j}\Delta z)\xi _{j}(z)\xi _{i}(z^{\prime })}{(\Delta z)^3}.
\label{exchange}
\end{equation}
The dependence on the slab width $d$ in Eqs.~(\ref{kinetic}), (\ref
{electrostatic}), and (\ref{exchange}) enters through the self-consistent KS
eigenvalues $\left( \varepsilon _{i}\right) $ and eigenfunctions $\left(\xi
_{i}(z)\right)$.

Alternatively, one can define the effective single-slab surface energies \cite{pitarke}

\begin{equation}
\sigma(d)=\frac{E(d)-E^{unif}(d)}{2A}  \label{surfener.2}
\end{equation}
and
\begin{equation}
\sigma_{l}(d)=\left[ E_{l}(d)-E^{unif}_{l}(d)\right] /(2A), 
\label{singleslab}
\end{equation}

where $E^{unif}_{l}(d)$ is the ground-state energy of a uniform slab of electron gas
of size $d$, and $l=K,\;el,\,x.$ \cite{formu.1}
Notice that Eq.~(\ref{surfener.2}) only
reproduces the surface-energy definition of Eq.~(\ref{surfener}) as
$d\to\infty$. However, for a correct extrapolation of finite-slab calculations
to the infinite-width limit,\cite{pitarke} here we calculate numerically the
three components of the surface energy from the single-slab
Eq.~(\ref{singleslab}). We have checked that the differences between surface energies obtained 
from Eq.~(\ref{surfener}) and Eq.~(\ref{surfener.2}) are quite small even for the narrowest 
slabs studied, and that both agree in the extrapolation towards the semi-infinite limit.

Being the ground-state density the basic ingredient of DFT, we found interesting 
to compare the differences between the different density profiles that we have obtained.
We exhibit in Figure~2 the self-consistent electron density profiles that we have
obtained within the $x$-only LDA and OEP schemes for $r_s=2.07$ and
$d=8\,\lambda_F$.\cite{note1} It is expected that the
amplitude of the difference between both densities diminishes as $z$
approaches the slab center, where both $n^{\text{LDA}}(z\rightarrow -\,d\,/\,2)$
and $n^{\text{OEP}}(z\rightarrow -\,d\,/\,2)$ should approach $\overline{n}$ as 
$d\rightarrow\infty .$ Fig.~2 shows that there are noticeable
differences between both densities: $n^{\text{LDA}}(z)$
extends further into the vacuum region than $n^{\text{OEP}}(z)$, which is a
result of the LDA orbitals being more extended or ``diffuse'' than
their OEP counterparts, and the amplitude of the Friedel oscillations near the
surface is larger for $n^{\text{OEP}}(z)$ than for $n^{\text{LDA}}(z)$. 
We have found the same behaviour for other values of $r_{s}$.

Figure~3 shows the results that we have obtained for the slab kinetic
surface energy, as a function of the slab width $d,$ for $r_{s}=2.07$. 
As in the case of the electron density, we have performed these calculations
within the $x$-only LDA and OEP schemes. In the LDA, the kinetic surface
energy $\sigma _{K}(d)$ (LDA) is obtained by introducing the $x$-only
self-consistent LDA eigenfunctions $\xi _{i}^{\text{LDA}}(z)$ and eigenvalues
$\varepsilon _{i}^{\text{LDA}}$ into the formally exact Eq.~(\ref{kinetic}). In
the OEP, the kinetic surface energy $\sigma _{K}(d)$ (OEP) is obtained by
using the same Eq.~(\ref{kinetic}) but with the LDA eigenfunctions and
eigenvalues replaced by their $x$-only OEP counterparts
$\xi_{i}^{\text{OEP}}(z)$ and $\varepsilon_{i}^{\text{OEP}}$. The strong
oscillations in both
$\sigma _{K}(d)$ (LDA) and $\sigma _{K}(d)$ (OEP) are the result of the
sequential filling of empty slab discrete levels as $d$ increases. Maxima in
$\sigma_{K}(d)$ correspond to the onset for the filling of a new slab discrete
level. For this particular case, and following the extrapolation procedure of
Ref.~\onlinecite{pitarke}, we have obtained the infinite-width extrapolated
surface energies $\sigma_K{\rm(\text{LDA}})=-\,4832\,{\rm erg/cm^2}$ (as reported in
Ref.~\onlinecite{pitarke}) and $\sigma_{K}{\rm(OEP)}=-\,4720\,{\rm erg/cm^2}.$ 

Figure~4 displays the results that we have obtained for the electrostatic
contribution to the surface energy, as a function of the slab width $d$ and
for $r_{s}=2.07$, again within the $x$-only LDA and OEP schemes. The
electrostatic surface energies $\sigma_{el}(d)$ (LDA) and $\sigma_{el}(d)$
(OEP) are obtained from Eq.~(\ref{electrostatic}) by using either the $x$-only
LDA electron density
$n^{\text{LDA}}(z)$ or the $x$-only OEP electron density $n^{\text{OEP}}(z)$, respectively.
In this case, the onset for the filling of a new slab discrete level is always
associated with a minimum. Following the extrapolation procedure of
Ref.~\onlinecite{pitarke}, we have obtained the infinite-width surface
energies indicated by arrows in Fig.~4: $\sigma_{el}{\rm(\text{LDA})}=1172\,{\rm
erg/cm^2}$ (as reported in Ref.~\onlinecite{pitarke}) and
$\sigma_{el}{\rm(\text{OEP})}=1103\,{\rm erg/cm^2}.$   

In Fig.~5, we show the results that we have obtained for the exact-exchange
contribution to the slab surface energy, as a function of the slab width $d$
and for $r_{s}=2.07$, again within the $x$-only LDA and OEP schemes. As in the
case of the kinetic and electrostatic surface energies, exact-exchange surface
energies $\sigma_{x}(d)$ (LDA) and $\sigma_{x}(d)$ (OEP) [both derived from
the formally exact Eq.~(\ref{exchange})] are obtained by using either the
$x$-only self-consistent LDA eigenfunctions $\xi _{i}^{\text{LDA}}(z)$ and
eigenvalues $\varepsilon _{i}^{\text{LDA}}$ or their $x$-only OEP counterparts
$\xi_{i}^{\text{OEP}}(z)$ and $\varepsilon_{i}^{\text{OEP}}$. For comparison,
we have also calculated standard LDA-exchange surface energies \cite{lang}

\begin{equation}
\sigma_{x}^{\text{LDA}}= \frac{1}{2}\int\limits_{-\infty}^\infty dz \,\, n^{\text{LDA}}(z)
\left\{\varepsilon_x^{unif}[n^{\text{LDA}}(z)]-
\varepsilon_x^{unif}(\bar n)\right\},
\label{ldax}
\end{equation}

where $\varepsilon_{x}^{unif}(n)$ is the exchange energy per particle  of a
uniform electron gas of density $n$:
$\varepsilon_x^{unif}(n)=-3e^2(3\pi^2n)^{1/3}/(4\pi)$, and
$n^{\text{LDA}}(z)$ represents the $x$-only LDA electron density.

All $\sigma_x(d)$ (LDA), $\sigma_x(d)$ (OEP), and $\sigma_x^{\text{LDA}}(d)$ exhibit the
characteristic oscillatory behaviour also shown by the other components of the
surface energy. As in the case of the electrostatic surface energy, the onset
for the filling of a new slab discrete level is associated with a minimum.
Fig.~5 shows that while the LDA [see Eq.~(\ref{ldax})] considerably
overestimates the exchange surface energy, which is a known result, the
exact-exchange surface energy is not very sensitive to the actual shape of the
single-particle orbitals and energies, i.e., to whether LDA or OEP orbitals
are used. Following the extrapolation procedure of Ref.~\onlinecite{pitarke},
we have obtained the infinite-width surface energies indicated by arrows in
Fig.~5: $\sigma_x^{\text{LDA}}=2767\,{\rm erg/cm^2}$,  $\sigma_x{\rm(
\text{LDA})}=2390\
{\rm erg/cm^2}$ (both as reported in Ref.~\onlinecite{pitarke}), and
$\sigma_{x}{\rm(OEP)}=2316\,{\rm erg/cm^2}.$

We have also computed kinetic, electrostatic, and exchange surface energies for
other values of the electron-density parameter $r_s$, and we have obtained the
infinite-width extrapolated results shown in Table I. A comparison of the LDA
and OEP calculations presented in Table I shows that (i) LDA orbitals being
more delocalized than the more realistic OEP orbitals, surface energies that
are based on the use of LDA orbitals are too large relative to those obtained
with the use of OEP orbitals, and (ii) the sum of kinetic, electrostatic, and
exchange surface energies are not very sensitive to whether LDA or OEP is used
in the evaluation of the single-particle KS eigenfunctions and eigenvalues    

\begin{table}
\caption{Infinite-width extrapolated results for exchange-only kinetic [$\sigma_{K}$ (LDA) and $\sigma _{K}$ (OEP)],
electrostatic [$\sigma _{el}$ (LDA) and $\sigma _{el}$ (OEP)],
and exchange [$\sigma_{x}^{\text{LDA}}$, $\sigma _{x}$ (LDA), and $\sigma _{x}$ (OEP)]
surface energies for different values or $r_{s}$. $\sigma$ (LDA) and
$\sigma$ (OEP) represent the sum of the corresponding exchange-only kinetic,
electrostatic, and exchange surface energies. Empty entries in $\sigma _{el}$ (OEP)
 for the two largest $r_s$ studied are due to the fact that the corresponding 
 magnitudes are so small that it is not possible obtain a reliable extrapolated value.
   Units are erg/cm$^2$.}

\begin{tabular}{|c|c|c|c|c|c|c|c|c|c|}
\hline
$r_{s}$ & $\sigma_{K}$ (LDA) & $\sigma_{K}$ (OEP) & $\sigma_{el}$ (LDA) & 
$\sigma_{el}$ (OEP) & $\sigma_{x}^{\text{LDA}}$ & $\sigma_{x}$ (LDA) &
$\sigma_{x}$ (OEP) & $\sigma$ (LDA) & $\sigma$ (OEP)  \\ 
\hline
2.00 & - 5707 & - 5579 & 1390 & 1317 & 3131 & 2726 & 2649 & -1591 & -1613 \\ 
\hline
2.07 & - 4832 & - 4720 & 1172 & 1103 & 2767 & 2390 & 2316 & -1270 & -1301 \\ 
\hline
3.00 & -770 & - 733 & 189 & 177 & 707 & 568 & 535 & -13 & -21 \\ \hline
4.00 & - 169 & - 155 & 49 & 48 & 243 & 180 & 161 & 60 & 54 \\ \hline
5.00 & -46 & - 39 & 19 & - & 105 & 71 & 59 & 44 & -  \\ \hline
6.00 & - 13 & - 9 & 9 & - & 52 & 32 & 23 & 28 & -  \\ \hline
\end{tabular}
\end{table}

\section{Work function}

The work function $W$ is the minimum work that must be done to remove an
electron from the metal at zero-temperature. In the context of DFT, the
rigorous expression for the work function for a slab of thickness $d$ is\cite
{zhang1} 
\begin{equation}
W(d)=V_{\text{KS}}(\infty )-\mu,
\label{workfunction}
\end{equation}
where $\mu$ is the chemical potential. We note that as we are considering an electron system
that is infinite in the $x$ - $y$ plane, electronic relaxation effects after
removal of one electron are infinitesimal. For a slab geometry, the work
function becomes size-dependent through the chemical potential 
$\mu (\bar n,d).$ We are imposing the
boundary condition $V_{\text{KS}}(\infty )=0$; accordingly,
$W(d)=\,-\,\mu >0.$
Besides, the {\it only} energy of the full KS spectrum which has a physical
significance is precisely the energy of the highest occupied level, which can
be identified with $\mu$.
\cite{schulte1} The work function for a slab with  $r_s=2.07$ and
$d=4\,\lambda _{F}$ is shown schematically in Fig.~1. For this particular case,
nine SDL are occupied and  $\mu$ is between the nineth and tenth SDL.

Now we focus on the slab-width dependence of the work function. Figure~6 shows
the result of the $x$-only calculations that we have performed within LDA and
OEP [$W^{\text{LDA}}$ and $W^{\text{OEP}}$] for $r_s=2.07$. The weakly oscillating $x$-only
$W^{\text{LDA}}(d)$ is equivalent to the slab-width dependent work function
reported by Schulte a long time ago.~\cite{schulte} As discussed by Schulte,
the
oscillations in $W^{\text{LDA}}(d)$ are the result of a combination of the
shift of the bottom of the slab potential well and an effective film thickness
shift, both effects suffering from an abrupt change each time the number of
occupied
SDL changes by one. The important point here, however, is the much stronger
oscillations found in our $W^{\text{OEP}}(d)$ calculations, whose explanation
is provided now with some detail.

First of all, we note that, strictly speaking, the OEP work function
$W^{\text{OEP}}(d)$ exhibits discontinuities
  of large size each time a new SDL becomes infinitesimally occupied. The first discontinuity in
Fig.~6 appears at the 1 SDL $\rightarrow $ 2 SDL transition (for
$d\lesssim\lambda_F/2$), the second discontinuity appears at the 2
SDL $\rightarrow $ 3 SDL transition (for $d\lesssim\lambda_F$),
and so on. In order of clarify the source of such a discontinuous behavior,
we have plotted in Fig.~7 the OEP exchange potential $V_x(z)$ for slightly
increasing values of the slab width $d,$ around the 6 SDL $\rightarrow $ 7
SDL transition. Each slab width $d$ is characterized by a ``filling factor'' of
the last occupied SDL, which is defined as follows 
\begin{equation}
\alpha _{m}\equiv \frac{\mu-\varepsilon _{m}}{%
\varepsilon _{m+1}-\varepsilon _{m}}.  \label{ff}
\end{equation}
Hence, $\alpha _{m}\rightarrow 0^{+}$
(implying $\mu \rightarrow \varepsilon _{m}^{+})$,
corresponds to an infinitesimally small filling of the last occupied SDL
$(i=m)$, while $\alpha _{m}\rightarrow 1^{-}$ corresponds to the threshold of
occupancy of the next SDL $(i=m+1).$ The key
point here is the dramatic change in $V_{x}(z)$ when passing from the slab
thickness corresponding to $\alpha_{6}=1^{-}$ to the infinitesimally thicker
slab corresponding to $\alpha _{7}=0^{+}\,(\sim 10^{-5}).$ The remaining
curves have been obtained for slab widths corresponding to the seventh SDL
being progressively occupied: As $\alpha_7$ increases from $0^+$ to $1^-$,
$V_{x}(z)$ approaches the form it had at $\alpha _{6}=1^{-},$ both in depth
and asymptotic behavior, the only difference being a lateral shift of
$V_{x}(z)$ to the right that is simply due to the larger value of
$d$.

Secondly, we note that the potential barrier that forms at the interface, right
after the jellium edge on the vacuum side of the surface, exhibits both
$V_{x,1}(z)$ and $V_{x,2}(z)$ contributions
[see Eq.~(\ref{oeppotentialdecomposition}], so the KLI approximation (which
sets $V_{x,2}(z)\equiv 0$) cannot be used for the analysis of the
characteristic discontinuous behavior of the work function. In all cases in
Fig. 7, $V_{x}(z\to\infty)\to 0.$ While this is clearly seen in the figure for
the curves corresponding to $\alpha_6=1^{-}$ and $\alpha_{7}=1^{-}$ [in which
case $k_F^m\sim 1/d$; see the asymptotics of Eq.~(\ref{epasymp})], it is not
evident at all for the set of potentials with small occupancies of the last
occupied level, i.e., $\alpha_7<<1$. In this case, $k_F^m<<1/d$ and the
asymptotic regime only takes place at $z$ coordinates that go to infinity (as
$\alpha_7\to 0^+$) far beyond the $z$ coordinates considered in Fig.~7. This
is the situation for $\alpha _{7}\simeq 10^{-5},\,10^{-4},$ and $10^{-3}.$ As
a final remark on this figure, it is important to realize that in the bulk and
near the interface the exchange potentials $V_{x}(z)$ corresponding to $\alpha
_{6}=1^{-}$ and $\alpha _{7}=0^{+}$ are simply related through a single
vertical (constant) shift. This property, which can be verified numerically
from Fig.~7, may also be derived analytically (see below). Finally, we note
that although we have restricted our discussion to the case of a particular
SDL transition, the same happens at every highest occupied $\rightarrow$
lowest unoccupied SDL transition.

With the aim of understanding how this discontinuous behavior of $V_{x}(z)$
versus the slab width explains the results of Fig.~6 for the work function
$W^{\text{OEP}}(d)$, we show in Fig.~8 the slab OEP electronic structure just
before occupation of the SDL \# 7 (left panel), that is, at the slab width
corresponding to $\alpha _{6}\rightarrow 1^{-},$ and just after occupation of
the SDL \# 7 (right panel), i.e., at the slab width corresponding to
$\alpha_{7}\rightarrow 0^{+}.$ We note that while the Hartree potential
approaches zero outside the surface exponentially and remains essentially
unaffected by the infinitesimal population of the SDL \#7 (compare left and
right panels of Fig.~8), the OEP exchange potential (and therefore
$V_{\text{KS}}(z)$ as well) suffers the abrupt jump explained in Fig.~7 which
induces in turn the corresponding abrupt jump in the Fermi level. The net
result in going from the left to the right panels of Fig.~8 is that the work
function $W^{\text{OEP}}(d)$ suffers an abrupt (discontinuous) decrease, as
the boundary condition $V_{\text{KS}}(\infty )=0$ is rigorously valid in both
cases. This discontinuous behavior of $W^{\text{OEP}}(d)$, shown schematically
in Fig.~8, represents precisely the origin of the jumps that are visible in
Fig.~6 at every threshold for SDL occupation. It is evident from Fig.~6 that
the size of the discontinuity decreases as $d$ increases.

Finally, we investigate the size of the discontinuities that are visible in
Fig.~6. For this, we rewrite the central OEP equation [as given by
Eq.~(\ref{oepcompact})] in the following way: 
\begin{equation}
\begin{array}{l}
\sum\limits_{i=1}^{m-1}(k_{F}^{i})^{2}\int\limits_{-\infty }^{\infty }\left[
V_{x}(z^{\prime };m)-u_{x}^{i}(z^{\prime };m)\right] G^{\text{KS}}_{i}(z,z^{\prime
})\,\varphi _{i}(z^{\prime },z)\,dz^{\prime }+ \\ 
(k_{F}^{m})^{2}\int\limits_{-\infty }^{\infty }\left[ V_{x}(z^{\prime
};m)-u_{x}^{i}(z^{\prime };m)\right] G^{\text{KS}}_{m}(z,z^{\prime })\,\varphi
_{m}(z^{\prime },z)\,dz^{\prime }+\text{c.c.}=0,
\end{array}
\label{oepsplitted}
\end{equation}
where $\varphi _{i}(z,z^{\prime })=\xi_{i}(z)^{*}\xi_{i} (z^{\prime }).$ In
writing Eq.~(\ref{oepsplitted}) the contribution of all the $m-1$ occupied
SDL's has been split from the contribution of the last occupied $(m)$ SDL. The
label $m$ in $V_{x}(z;m)$ and $u_{x}^{i}(z;m)$ has been introduced in order to
emphasize that they are solutions of a system with $m$ occupied SDL's.

Let us now define a distance $Z,$ such that for $z>Z$ the electron density is
dominated by the contribution of the last occupied ($m$) SDL, which is the one
with the slowest decay. Eq.~(\ref{density}) clearly shows that
$Z\rightarrow \infty$ when $k_{F}^{m}\rightarrow 0,$ which is the case whenever
$\alpha_m\to 0^+$, i.e., whenever the filling of the last occupied SDL is
infinitesimally small. We consider the following trial solution of
Eq.~(\ref{oepsplitted}): 
\begin{equation}
V_{x}(z;m)=V_{x}(z;m-1)+C_{x}(m),  \label{oepsolution}
\end{equation}
for $z<Z$ and $k_{F}^{m}\rightarrow 0,$ with $C_{x}(m)$ being a constant which
depends on the last occupied SDL$.$ Introducing this trial solution into
Eq.~(\ref{oepsplitted}), we obtain 
\begin{equation}
\begin{array}{l}
\sum\limits_{i=1}^{m-1}(k_{F}^{i})^{2}\int\limits_{-\infty }^{\infty }\left[
V_{x}(z^{\prime };m-1)+C_{x}(m)-u_{x}^{i}(z^{\prime };m)\right]
G^{\text{KS}}_{i}(z,z^{\prime })\,\varphi _{i}(z^{\prime },z)\,dz^{\prime }+ \\ 
(k_{F}^{m})^{2}\int\limits_{-\infty }^{\infty }\left[ V_{x}(z^{\prime
};m-1)+C_{x}(m)-u_{x}^{i}(z^{\prime };m)\right] G^{\text{KS}}_{m}(z,z^{\prime
})\,\varphi _{m}(z^{\prime },z)\,dz^{\prime }+\text{c.c.}=0.
\label{lastoep}
\end{array}
\end{equation}
In the limit $k_{F}^{m}\rightarrow 0,$ the second-line contribution of
Eq.~(\ref{lastoep}) is arbitrarily small; also, the KS wave-functions
$\xi _{i}(z)$ and eigenvalue
differences (denominators) entering $G^{\text{KS}}_{i}(z,z^{\prime })$ should be extremely
similar for the slab width corresponding to $m-1$ occupied levels and
$\alpha_{m-1}\rightarrow 1^{-},$ and the slab width corresponding to $m$
occupied levels and $\alpha _{m}\rightarrow 0^{+}.$ Therefore, an inspection
of Eq.~(\ref{odp}) leads us, using similar arguments, to the conclusion that
$u_{x}^{i}(z;m)\rightarrow $ $u_{x}^{i}(z;m-1),$ for all
$i<m,$ $z<Z,$ and $k_{F}^{m}\rightarrow 0.$ Under these conditions, the first
line of Eq.~(\ref{lastoep}) reverts to the OEP equation for a slab width
corresponding to $m-1$ occupied states, and the proposal of
Eq.~(\ref{oepsolution}) is proved. Considering now that
$C_{x}(m)=V_{x}(z;m)-V_{x}(z;m-1),$ and taking the expectation value at the
last occupied
state
($m-1$) of $m-1$ system, we find 
\begin{equation}
C_{x}(m)=\overline{V}\,_{x}^{m-1}(m)-\overline{V}\,_{x}^{m-1}(m-1).
\label{constant}
\end{equation}
Now, for the $m-1$ system we can use the boundary condition $\overline{V}%
\,_{x}^{m-1}(m-1)=\overline{u}\,_{x}^{m-1}(m-1),$ and once again,
approximate
$\overline{u}\,_{x}^{m-1}(m-1)\simeq \overline{u}\,_{x}^{m-1}(m),$ yielding 
\begin{equation}
C_{x}(m)=\overline{V}\,_{x}^{m-1}(m)-\overline{u}\,_{x}^{m-1}(m),
\label{constant1}
\end{equation}
which has the nice feature that both the exchange potential $V_x$ and the
orbital-dependent exchange potential $u_x$ are referred to the $m$ system.
For the $m$ system $\overline{V}\,_{x}^{m}(m)=\overline{u}\,_{x}^{m}(m),$ which
does not prevent the constant $C_x(m)$ from being nonzero (as shown in Fig.~7)
since the KS orbitals $\xi _{m-1}(z)$ and $\xi _{m}(z)$ are different. As the
slab width increases, $m$ also increases and the difference between
$\xi _{m-1}(z)$ and $\xi _{m}(z)$ decreases, thereby leading to the expectation
that $C_{x}(m)\rightarrow 0$ as $d\rightarrow \infty.$ This is explicitly
shown in Fig.~9. While this analysis explains why $C_{x}(m)\neq 0$ for any
finite
$m,$ it does not gives a hint about its sign; Fig.~9 shows, however, that
$C_{x}(m)$ is positive for all $m.$ This positive jump in $V_{x}(z)$ is
exchange driven: at each threshold width for the occupation of a new level, a
barrier appears against the occupancy of an empty SDL. This is due to the fact
that intra-SDL exchange is stronger than inter-SDL exchange. As a consequence,
the slab gains exchange energy by restricting new SDL occupancies. On the
other
hand, correlation induces in general a negative jump in $V_{c}(z),$ so the
net jump in $V_{xc}(z)$ depends on the relative weigth of exchange and
correlation for each particular system.~\cite{rigamonti3} 

Finally, we have observed numerically that the average of the OEP work
functions for slab widths corresponding to $\alpha _{m-1}\rightarrow 1^{-}$
and
$\alpha _{m}\rightarrow 0^{+}$ remains the same (within error bars) for all the
$m$ values that we have considered. Hence, we have taken the infinite-width
extrapolated work function to be simply that average. Table II exhibits the
infinite-width $x$-only LDA and OEP work functions that we have obtained in
this way for various values of the electron-density parameter $r_s.$ OEP work functions
are slightly and sistematically smaller than their LDA counterparts.\\
    
\begin{table}
\caption{Infinite-width extrapolated $x$-only LDA and OEP work functions for
various values of $r_s$. Units are eV.}  
\begin{tabular}{|c|c|c|c|c|c|c|}
\hline
$r_{s}$ & 2.00 & 2.07 & 3.00 & 4.00 & 5.00 & 6.00  \\ 
\hline
$W^{\text{LDA}}$& 2.82 & 2.80 & 2.50 & 2.15 & 1.86 & 1.62 \\ 
\hline
$W^{\text{OEP}}$& 2.64 & 2.63 & 2.49 & 2.11 & 1.84 & 1.61 \\
\hline
\end{tabular}
\end{table}

\section{Conclusions}

We have reported benchmark exact-exchange self-consistent calculations of the
KS potential, surface energy, and work function of jellium slabs in the
framework of the OEP scheme. Special emphasis has been put into the
asymptotical behaviour of the exact-exchange KS potential far into the vacuum
and the large quantum size effects that are present in the slab-width
dependence of the surface energy and work function.

We have performed a detailed analysis of the asymptotics of the exact-exchange
KS potential far into the vacuum \cite{horowitz}, showing that at a distance $z$ that is
larger than the slab thickness the exact-exchange potential takes an
image-like form: $V_{x}(z\rightarrow \infty )\rightarrow -e^{2}\,/\,z,$ but
with a coefficient that differs from that of the classical image potential
$V_{im}(z)=-e^2/4z$. Although this result has been obtained in the $x$-only
approximation, it is also true in the presence of correlation due to the
separability of the basic OEP equations in their basic exchange and
correlation components.

The OEP kinetic, electrostatic, and exchange contributions to the surface
energy of jellium slabs have been obtained as a function of the slab width $d$
and for a set of electron densities characterized by the parameter $r_s$. We
have shown that these components of the surface energy are all oscillating
functions of $d,$ with the oscillating period being $\approx\lambda_{F}/2.$ By
a suitable extrapolation procedure, we have found the values of the different
components of the surface energy of a semi-infinite jellium. We have compared
our OEP surface energies with those obtained from the same formally exact
expressions [see Eqs.~(\ref{kinetic})-(\ref{exchange})] but using
single-particle LDA wave functions and energies; we have found small
differences between these OEP and LDA surface energies, which appear as a
consequence of the LDA orbitals being slightly more delocalized (diffuse) than
their more realistic OEP counterparts.

Finally, we have performed $x$-only OEP calculations of the work function of
jellium slabs, again as a function of the slab width $d$. We have found that
the OEP work function exhibits large quantum size effects that are absent in
the LDA and which reflect the intrinsic derivative discontinuity of the exact
KS potential. The amplitude of this discontinuity diminishes as the slab width
increases, and becomes arbitrarily small as $d\to\infty$, i.e, in the case of
a semi-infinite system. This has been proved both analytically and
numerically. We also note that although the precise value of the $x$-only OEP
work functions reported here would change with the inclusion of correlation,
the exact slab work function is expected to exhibit the large quantum size
effects and discontinuities observed in the present work, barring possible
accidental cancellations of exchange-driven and correlation-driven
contributions to the total discontinuity. The presence of
these large discontinuities in the $x$-only OEP slab work function (and
presumably also in the actual work function that includes correlation)
highlights the potential danger in which can be incurred by performing
elaborated calculations for a restricted set of slab sizes without
performing a suitable and reliable extrapolation towards the semi-infinite
case.

In summary, we expect that the {\it benchmark} exact-exchange OEP calculations
reported here for jellium slabs will serve as motivation and as a starting
point for the development of more realistic approximations for the exchange-correlation
 energy functional of jellium and real surfaces.

\section{Acknowledgments}

C.M.H wishes to acknowledge the financial support received
from CONICET of Argentina, through a Postdoctoral Fellowship. 
J.M.P. acknowledges
partial support by the University of the Basque Country, the Basque
Unibertsitate eta Ikerketa Saila, the Spanish Ministerio de Educaci\'on y
Ciencia (Grants No. FIS2006-01343 and CSD2006-53), and the EC 6th framework
Network of Excellence NANOQUANTA (Grant No. NMP4-CT-2004-500198).
C.R.P. was supported by the European Community through a Marie Curie IIF (MIF1-CT-2006-040222).

$^*$ Permanent address: Centro At\'{o}mico Bariloche and Instituto Balseiro, 8400 S. C. de
Bariloche, R\'{\i }o Negro, Argentina.

\newpage

{\bf FIGURE CAPTIONS.}\\

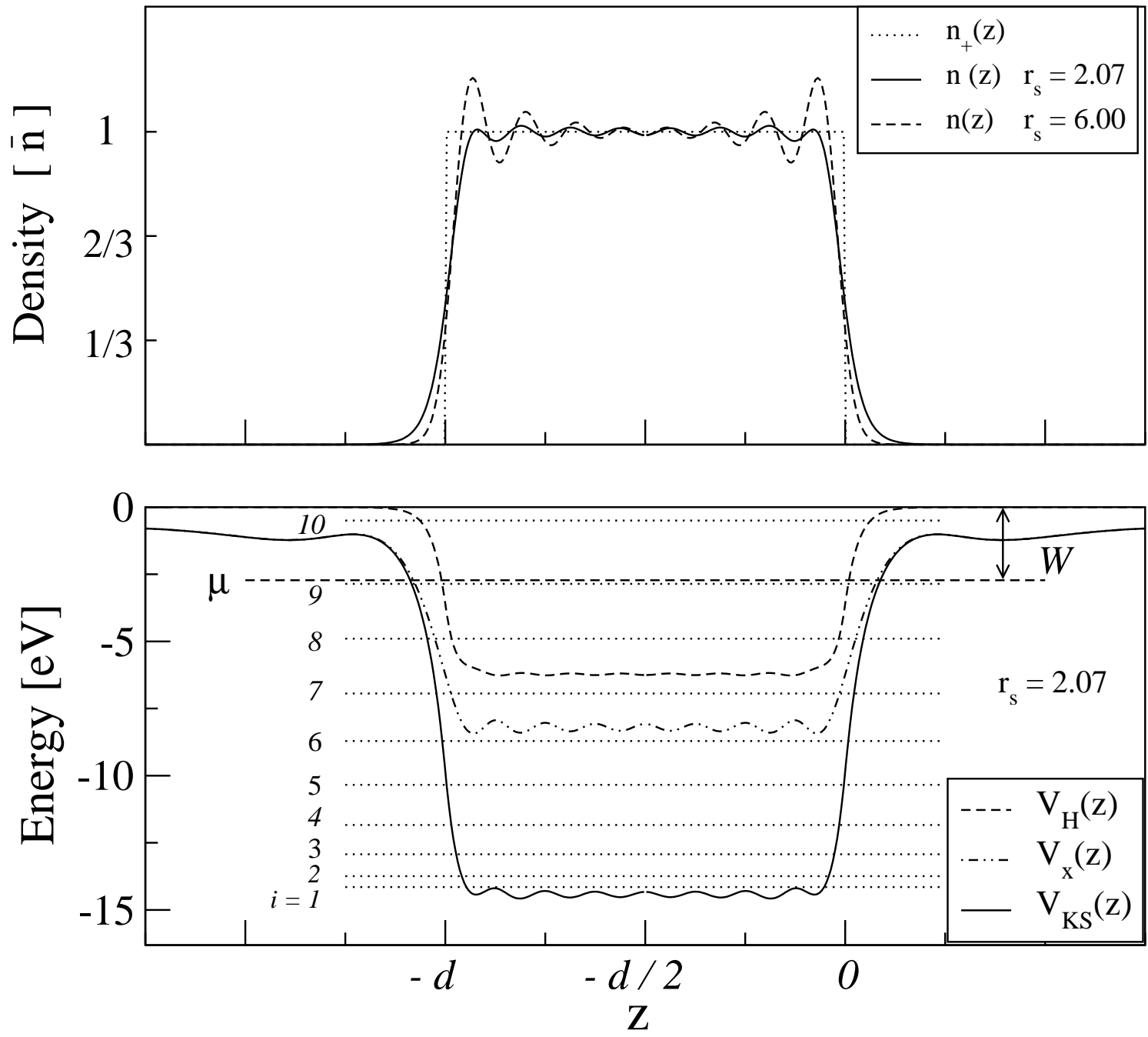
Figure 1. Main features of the jellium-slab model of metal surfaces. Top
panel: normalized jellium density $(n_{+}(z)),$ and the self-consistent OEP electron density
$n(z)$ for two different values of the electron-density parameter $r_s$. Lower
panel: OEP Hartree, exchange, and Kohn-Sham potentials for $r_{s}=2.07.$
Dotted lines denotes KS eigenvalues, $\varepsilon _{F}$ is the Fermi energy,
and $W$ is the work function. $d=4\;\lambda _{F}.$\\

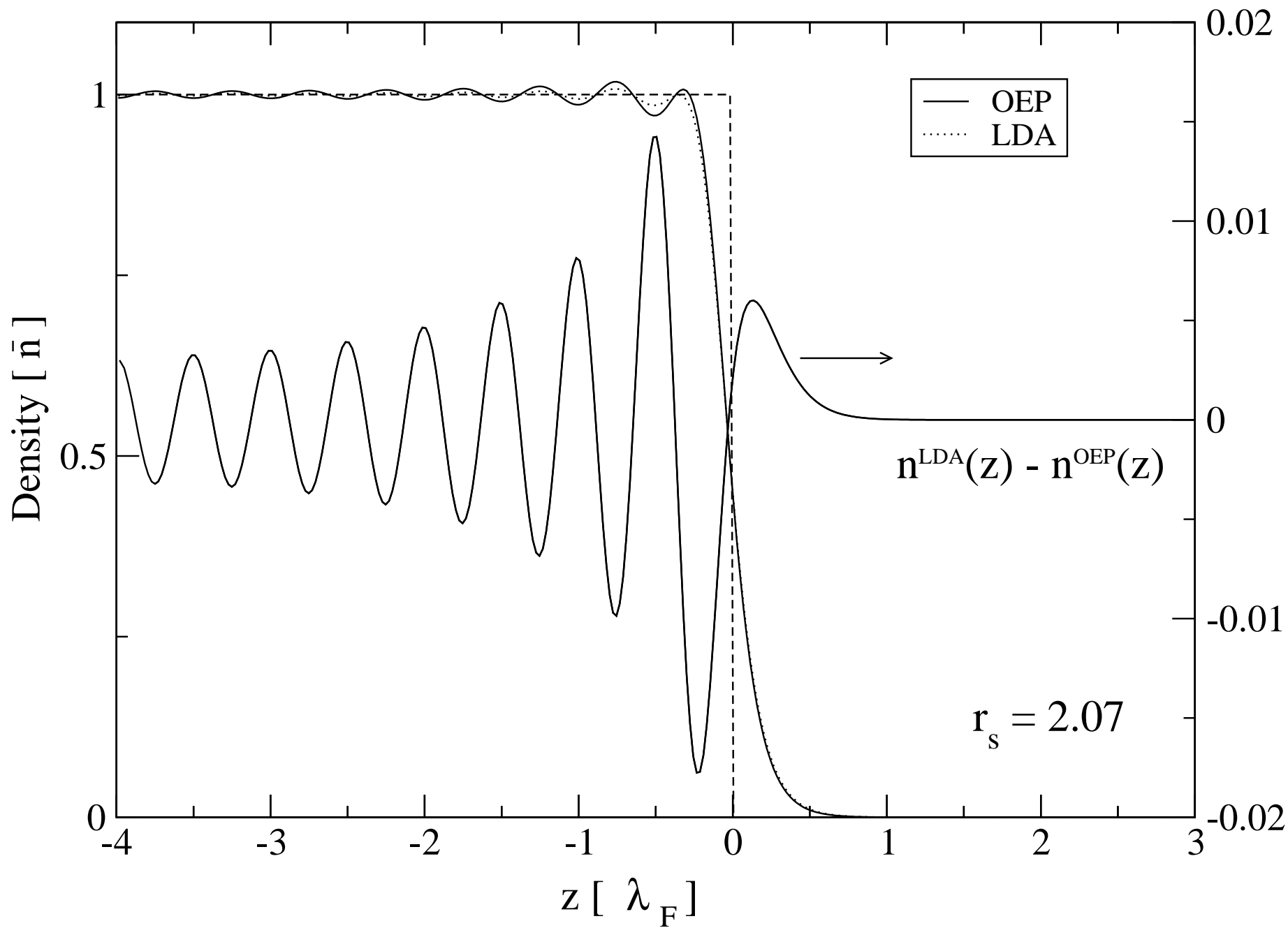
Figure 2. LDA and OEP self-consistent electron densities and its difference for
$d=8\,\lambda _{F}$ and $r_{s}=2.07$. Note that $n^{\text{LDA}}(z)$ is slightly more diffuse 
than $n^{\text{OEP}}(z)$, as $n^{\text{LDA}}(z) - n^{\text{OEP}}(z) >0$ for $z$ outside  
the jellium edge (in the vacuum).\\

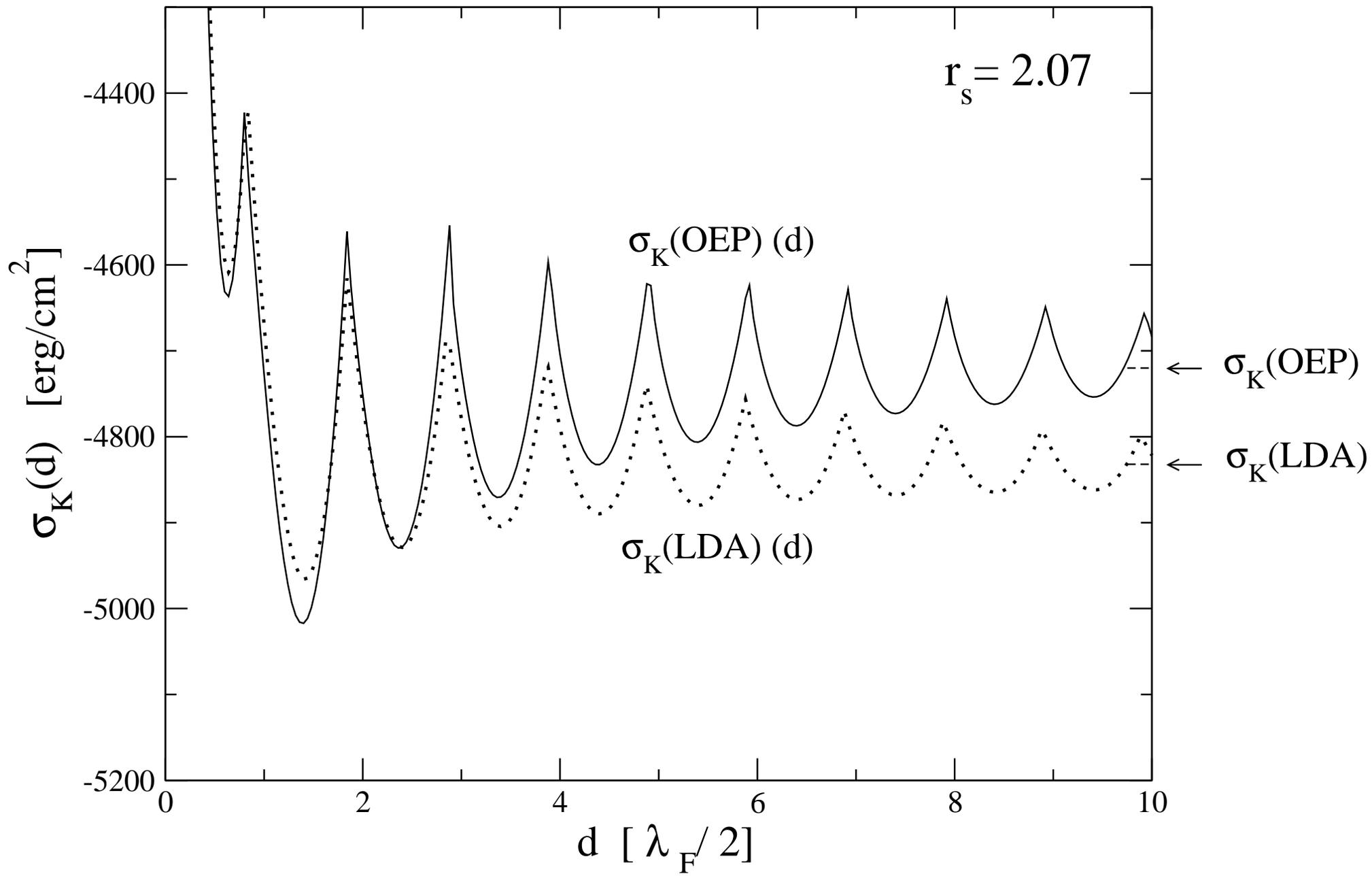
Figure 3. Kinetic surface energy, as a function of slab width $d,$ for $
r_{s}=2.07,$ from Eq.~(\ref{singleslab}), with $l=K$.
 Full line, OEP results; dotted line, LDA results. The two
arrows on the right denote the extrapolated asymptotic values $\sigma _{K}\,
 (\text{OEP}) \rightarrow -\,4720$ erg/cm$^{2},$ $\sigma _{K}\,(\text{LDA}) 
\rightarrow -\,4832$ erg/cm$^{2}.$\\

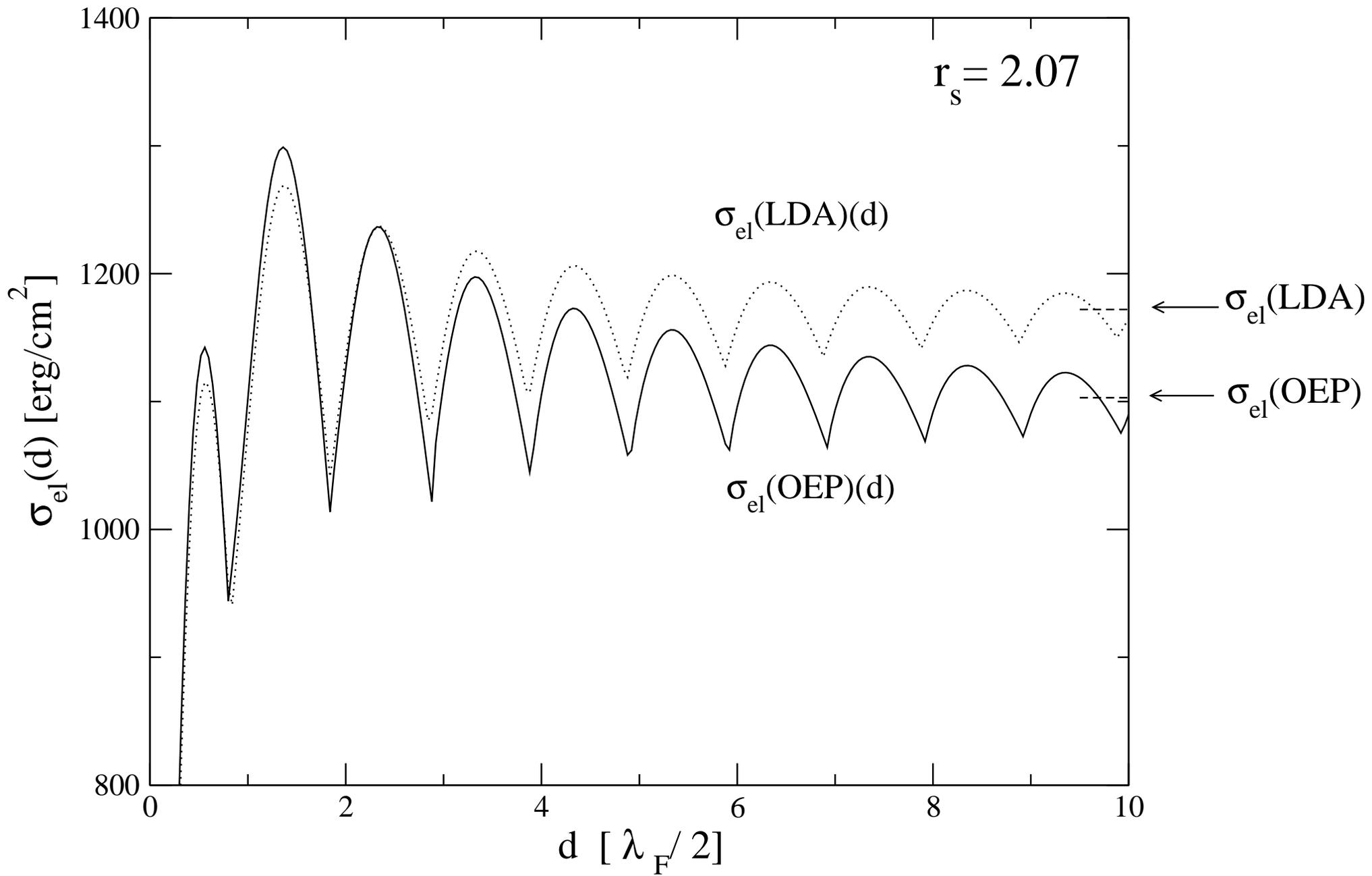
Figure 4. Electrostatic surface energy, as a function of slab width $d,$ for $%
r_{s}=2.07,$ from Eq.~(\ref{singleslab}), with $l=el$. Full line, OEP results; 
dotted line, LDA results. The two
arrows on the right denote the extrapolated asymptotic values 
$\sigma _{el}\,(\text{OEP})\rightarrow 1103$ erg/cm$^{2},$ $\sigma _{el}\,(\text{LDA})
\rightarrow 1172$ erg/cm$^{2}.$\\

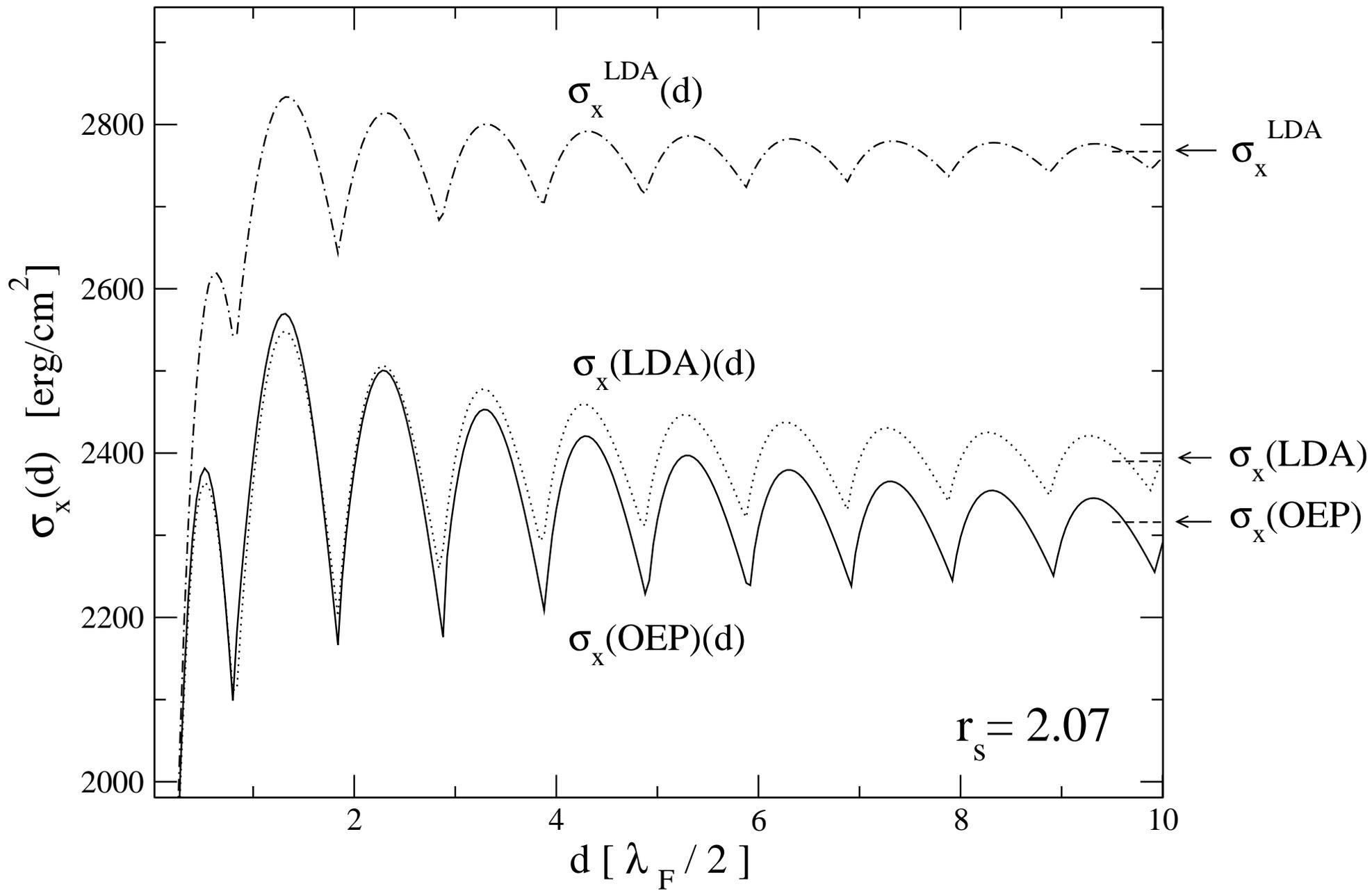
Figure 5. Exchange surface energy, as a function of slab width $d,$ for $
r_{s}=2.07,$ from Eq.~(\ref{singleslab}), with $l=x$. Full line, OEP results; 
dotted line, LDA results; dash-dotted
line, standard LDA-exchange results. The three arrows on the right denote the
extrapolated asymptotic values $\sigma _{x}(\text{OEP}) \rightarrow
2316$ erg/cm$^{2},$ $\sigma _{x}(\text{LDA}) \rightarrow 2390$
erg/cm$^{2}$, $\sigma _{x}^{\text{LDA}} \rightarrow 2767$
erg/cm$^{2}.$\\

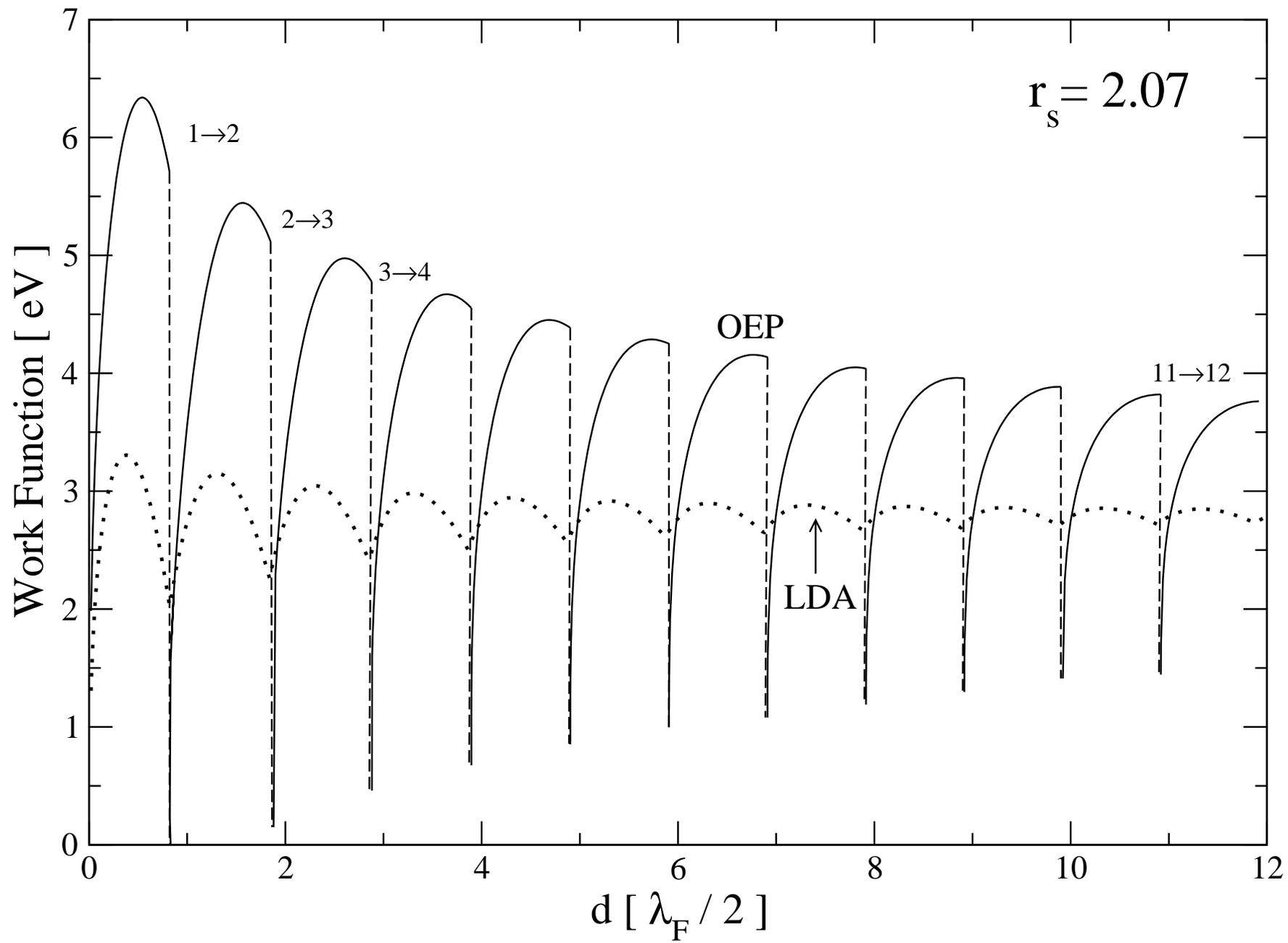
Figure 6. Slab work function versus slab width $d,$ for $r_{s}=2.07.$ Full
line, OEP result; dotted line, LDA result. Occupation events corresponding
to transitions from a slab with $m$ occupied SDL towards $m+1$ occupied SDL
are denoted as $m\rightarrow m+1.$\\

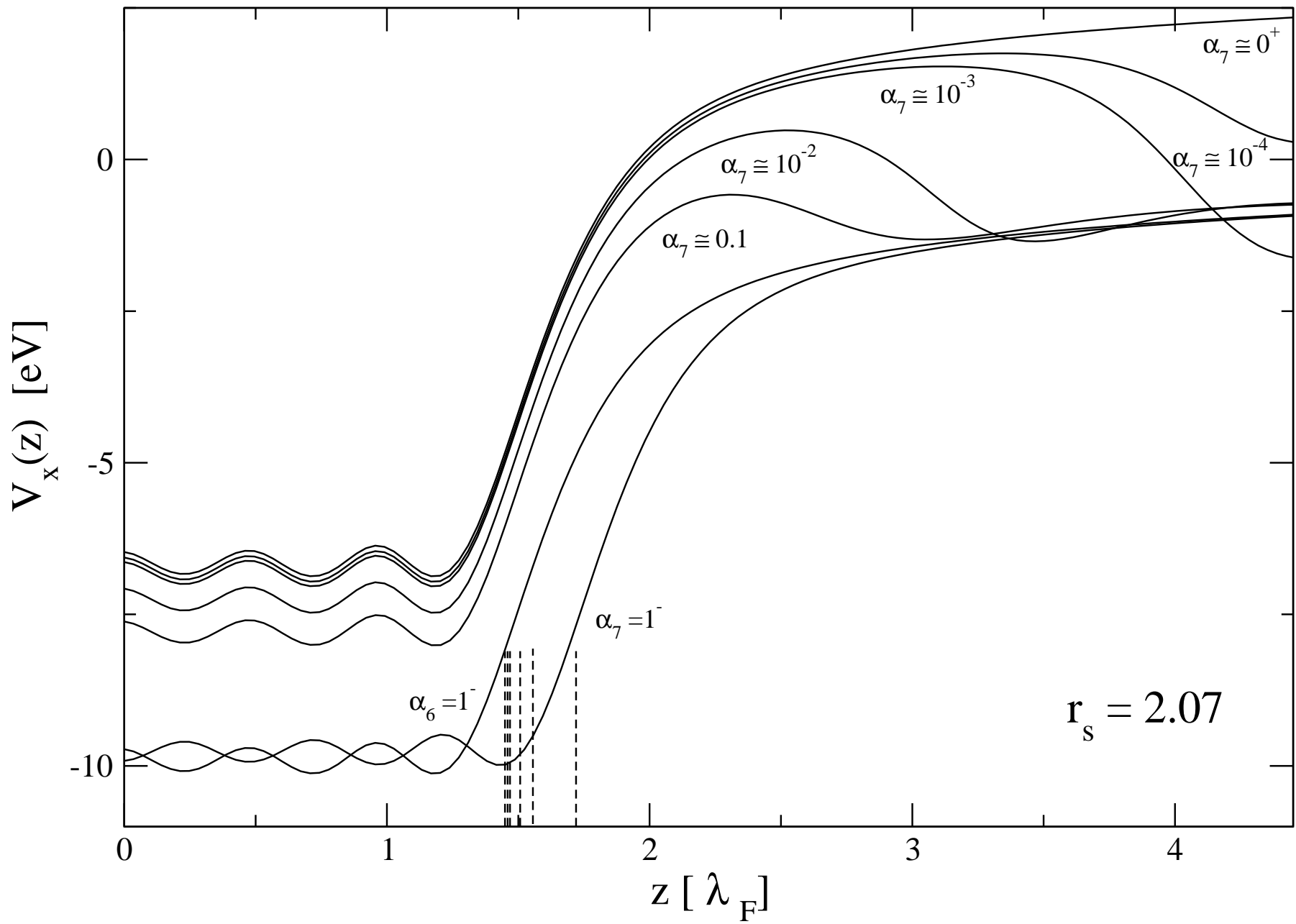
Figure 7. Self-consistent OEP exchange potential, around the $6\rightarrow 7$
SDL transition, for $r_{s}=2.07.$ The origin of coordinate $z$ for each slab
has been taken at the slab center. The position of the right slab edge has been indicated by a vertical dashed line for each case.\\

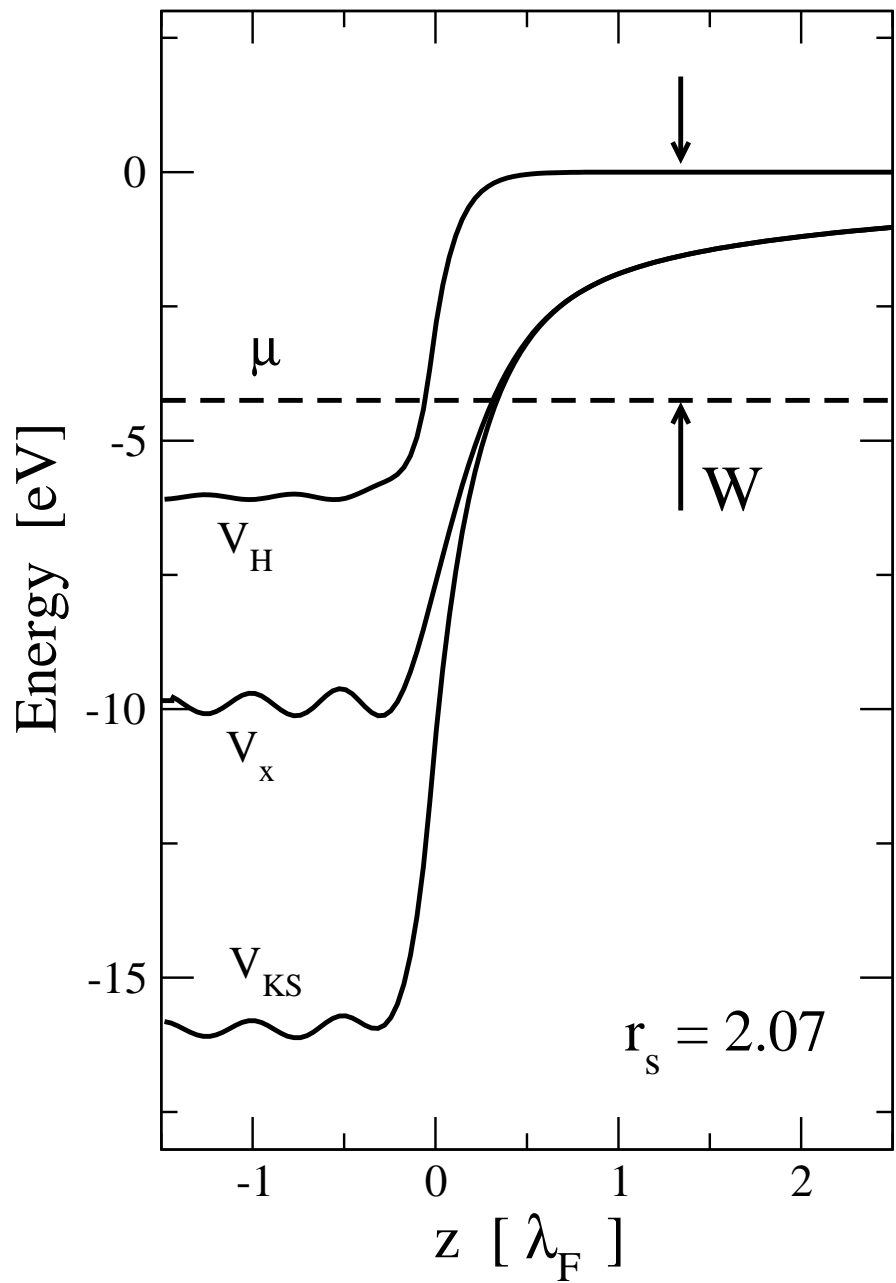
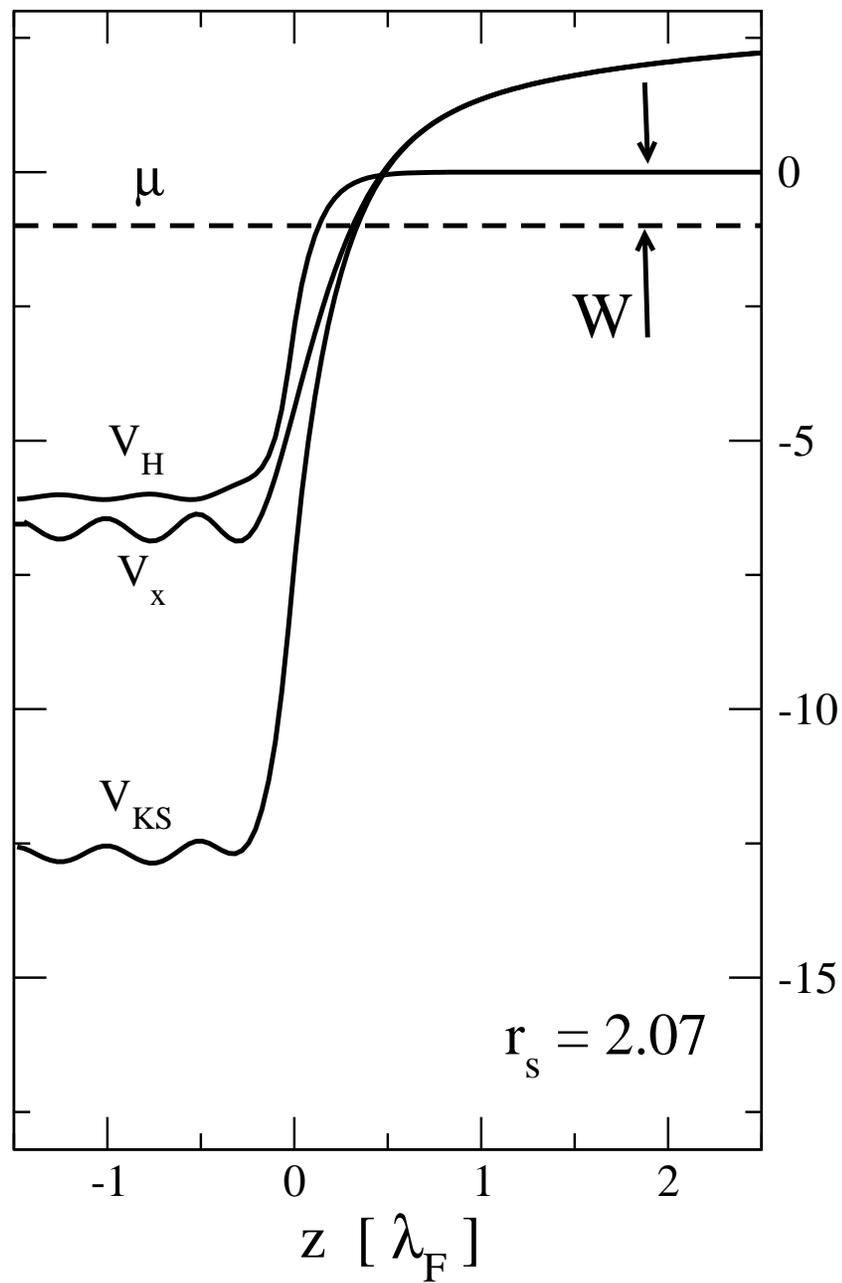
Figure 8. Left: electronic structure of the slab for $\alpha _{6}=1^{-}.$
Right: electronic structure of the slab for $\alpha _{7}=0^{+}.$ The work
function $W$ jumps discontinously from its left large value towards the
smaller right value. Slab edge is at $z=0$.\\

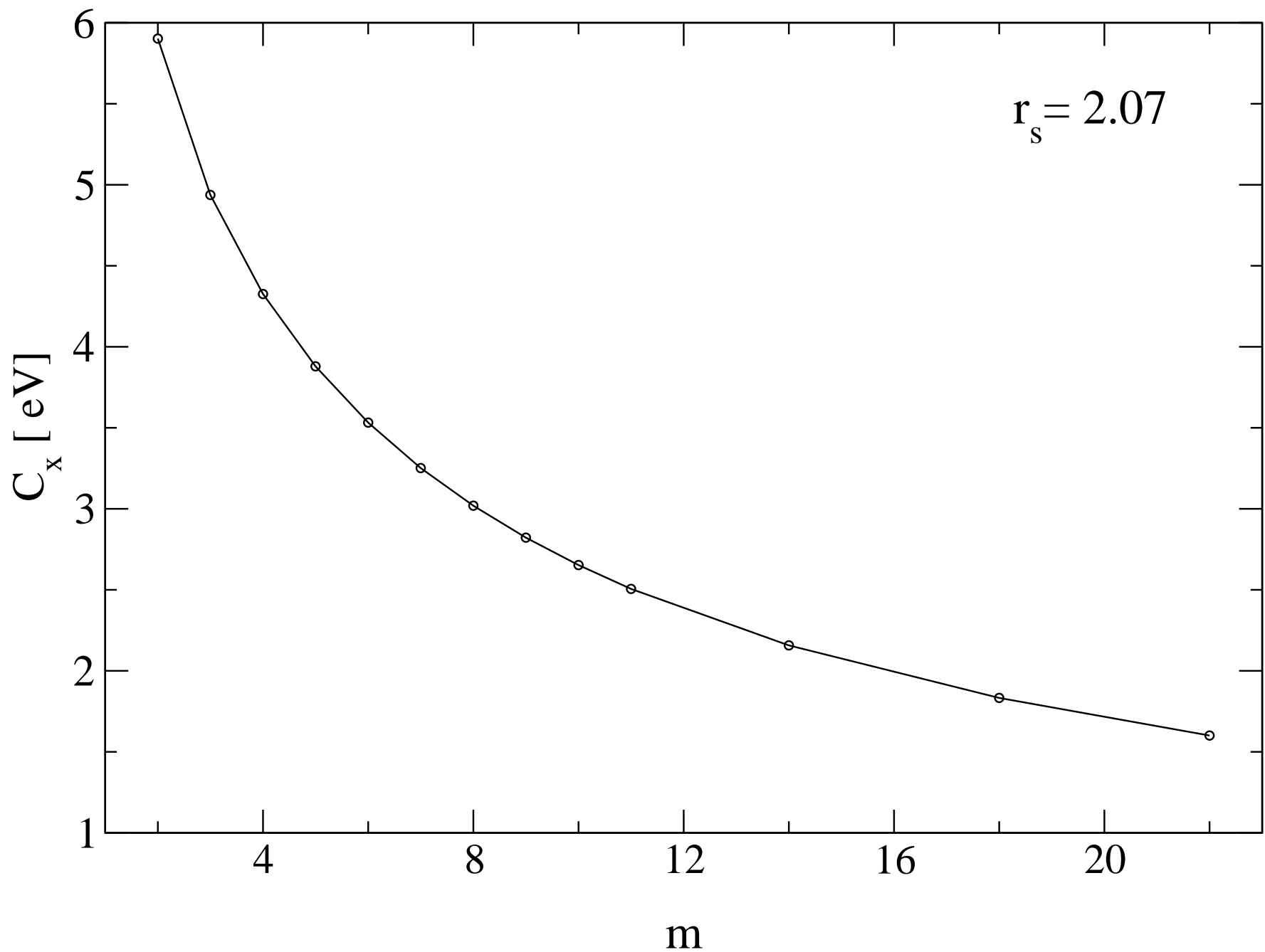
Figure 9. Exchange-driven discontinuity $C_{x}(m)$ for increasing number of
occupied slab levels, as follows from Eq. \ref{constant1}.\\


\begin{references}


\bibitem{lang}  N. D. Lang and W. Kohn, Phys. Rev. B {\bf 1}, 4555 (1970).

\bibitem{reviewlang}  N. D. Lang in {\it Solid State Physics}, edited by H.
Eirenreich, F. Seitz, and D. Turnball (Academic, New York, 1973), Vol. 28,
p. 225.

\bibitem{parr}  R. G. Parr and W. Yang, {\it Density Functional Theory of
Atoms and Molecules} (Oxford University Press, New York, 1989).

\bibitem{dreizler}  R. M. Dreizler and E. K. U. Gross, {\it Density
Functional Theory: An Approach to the Quantum Many-Body Problem}
(Springer-Verlag, Heidelberg, 1990).

\bibitem{ks}  W. Kohn and L. J. Sham, Phys. Rev. {\bf 140}, A1133 (1965).

\bibitem{nekovee}  M. Nekovee and J. M Pitarke, Computer Phys. Comm. {\bf 137%
}, 123 (2001).

\bibitem{zhang}  Z. Y. Zhang, D. C. Langreth, and J. P. Perdew, Phys. Rev. B 
{\bf 41}, 5674 (1990).

\bibitem{constantin}  L. A. Constantin, J. P. Perdew, and J. Tao, Phys. Rev.
B {\bf 73}, 205104 (2006).

\bibitem{kkq}  E. Krotscheck, W. Kohn, and G.-X. Qian, Phys. Rev. B {\bf 32}%
, 5693 (1985).

\bibitem{kk}  E. Krotscheck, W. Kohn, Phys. Rev. Lett. {\bf 57}, 862 (1986).

\bibitem{eguiluz}  A. G. Eguiluz, M. Heinrichsmeier, A. Fleszar, and W.
Hanke, Phys. Rev. Lett. {\bf 68}, 1359 (1992).

\bibitem{li}  X. P. Li, R. J. Needs, R. M. Martin, and D. M. Ceperley, Phys.
Rev. B {\bf 45}, 6124 (1992).

\bibitem{acioli}  P. H. Acioli and D. M. Ceperley, Phys. Rev. B {\bf 54},
17199 (1996).

\bibitem{wood}  B. Wood, N. D. M. Hine, W. M. C. Foulkes, and P. Garc\'{i}%
a-Gonz\'{a}lez, Phys. Rev. B {\bf 76}, 035403 (2007).

\bibitem{istls} L. A. Constantin, J. M. Pitarke, J. F. Dobson, A. Garc\'\i
a-Lekue, and J. P. Perdew, Phys. Rev. Lett. {\bf 100}, 036401 (2008).

\bibitem{grabo}  T. Grabo, J. Kreibich, S. Kurth, and E. K. U. Gross, in 
{\it Strong Coulomb Interactions in Electronic Structure Calculations:
Beyond the Local Density Approximation}, edited by V. I. Anisimov (Gordon
and Breach, Amsterdam, 2000).

\bibitem{engel}  E. Engel, in {\it A Primer in Density Functional Theory},
edited by C. Fiolhais, F. Nogueira, and M. A. L. Marques, Lecture Notes in
Physics, Vol. 620 (Springer, Berlin, 2003), p. 56.

\bibitem{talman}  J. D. Talman and W. F. Shadwick, Phys. Rev. A {\bf 14}, 36
(1976).

\bibitem{perdew}  J. P. Perdew and A. Zunger, Phys. Rev. B {\bf 23}, 5048
(1981).

\bibitem{dellasala}  F. Della Sala, and A. G\"{o}rling, Phys. Rev. Lett. 
{\bf 89}, 033003 (2002); J. Chem. Phys. {\bf 116}, 5374 (2002).

\bibitem{bylander}  D. M. Bylander and L. Kleinman, Phys. Rev. Lett. {\bf 74}%
, 3660 (1995); M. St\"{a}dele, J. A. Majewski, P. Vogl, and A. G\"{o}rling, 
{\it ibid.} {\bf 79}, 2089 (1997); Y.-H. Kim and A. Gorling, {\it ibid.} 
{\bf 89}, 096402 (2002).

\bibitem{vangisbergen}  S. J. A. van Gisbergen, P. R. T. Schipper, O. V.
Gritsenko, E. J. Baerends, J. G. Snijders, B. Champagne, and B. Kirtman,
Phys. Rev. Lett. {\bf 83}, 694 (1999).

\bibitem{kriebich}  T. Kriebich, S. Kurth, T. Grabo, and E. K. U. Gross,
Adv. Quantum Chem. {\bf 33}, 31 (1999).

\bibitem{kli}  J. B. Krieger, Y. Li, and G. J. Iafrate, Phys. Rev. A {\bf 45}%
, 101 (1992).

\bibitem{pplb}  J. P. Perdew, R. G. Parr, M. Levy, and J. L. Balduz, Phys.
Rev. Lett. {\bf 49}, 1691 (1982).

\bibitem{pl}  J. P. Perdew and M. Levy, Phys. Rev. Lett. {\bf 51}, 1884
(1983).

\bibitem{p}  J. P. Perdew, NATO ASI Ser., Ser. B {\bf 123}, 265 (1985).

\bibitem{tozer}  D. J. Tozer and N. C. Handy, J. Chem. Phys. {\bf 108}, 2545
(1998).

\bibitem{kim}  Y.-H. Kim, I.-H. Lee, S. Nagaraja, J. P. Leburton, R. Q.
Hood, and R. M. Martin, Phys. Rev. B {\bf 61}, 5202 (2000); P. Garc\'{i}%
a-Gonz\'{a}lez, ibid. {\bf 62}, 2321 (2000); L. Pollak and J. P. Perdew, J.
Phys.: Condens. Matter {\bf 12}, 1239 (2000); P. Garc\'{i}a-Gonz\'{a}lez and
R. W. Godby, Phys. Rev. Lett. {\bf 88}, 056406 (2002).

\bibitem{note0} The electron-density parameter $r_s$ is defined as the radius
of a sphere containing on average one electron, i.e.,
$r_s=(3/4\pi\bar n a_0^3)^{1/3}$. A convenient length unit for the present system is
$\lambda _{F}=(32\pi ^{2}/9)^{1/3}\;r_{s}a_{0} \approx 3.274\; r_{s}a_{0} $.

\bibitem{note0a} We have found that two infinite barriers located 
at $2 \lambda _{F}$ from each jellium edge are enough for this purpose in our system.

\bibitem{hartree.a} Note that we have included in our Hartree potential the contribution 
which comes 
from the uniform jellium background (proportional to $n_{+}$). Alternatively, 
this contribution may be denoted separately as the "external potential".

\bibitem{dens.aa} Note that dimensions of Eq. (6) are (length)$^{-3}$, as correspond 
to our 3D system. However, and for the particular slab geometry, the number density 
only depends on one spatial coordinate ($z$).

\bibitem{factor.a} Due to the imposed translational invariance in the $x$, $y$ plane, 
functional derivatives in this work are conveniently defined as 
$\delta f= \int [\delta f[g]/\delta g(z)]\delta g(z) dz$, where $\delta g(z)$ represents a 
uniform variation of the function $g({\bf r})$ in the plane ${\bf r}=z$. This is the origin
of the factor $A^{-1}$ in Eq. \ref{xcpotential}.  


\bibitem{exchangefunctional} We assume that the $xc$ energy functional does not
depend on the unoccupied eigenfunctions and eigenvalues of the KS equation,
which is only known to be true in the case of the $x$-only energy
functional. In general, the sum over the index $i$ should run over the whole
KS spectrum.\cite{rigamonti2,rigamonti3} Also, for simplicity, we do not consider the 
possibility that $E_{xc}$ depends on the Kohn-Sham eigenvalues. Once again, this is only 
known to be true in the case of the x-only energy functional, for fixed particle number. 
This is however, the only case to which all the numerical calculations presented below 
apply.  

\bibitem{ux.explicito} In the x-only version it can be written as Eq. \ref{odp}.

\bibitem{rigamonti2}  S. Rigamonti and C. R. Proetto, Phys. Rev. B {\bf 73},
235319 (2006).

\bibitem{rigamonti3}  S. Rigamonti and C. R. Proetto, Phys. Rev. Lett. {\bf %
98}, 066806 (2007).

\bibitem{kummel}  S. Kummel and J. P. Perdew, Phys. Rev. Lett. {\bf 90},
043004 (2003); {\it ibid}, Phys. Rev. B {\bf 68}, 035103 (2003).

\bibitem{rigamonti1}  S. Rigamonti, C. R. Proetto, and F. A. Reboredo,
Europhys. Lett. {\bf 70}, 116 (2005).

\bibitem{horowitz}  C. M. Horowitz, C. R. Proetto, and S. Rigamonti, Phys.
Rev. Lett. {\bf 97}, 026802 (2006). Note that there is a sign of difference
between the definition of the shifts in this and the present work.

\bibitem{kli.2} Note that the equivalence is only valid for this specific case 
(shifts identically zero). As soon as the shifts are different from 
zero, $V_{xc,1}(z)$ and $V_{xc}^{KLI}(z)$ will be not equal, due to the 
presence of $V_{xc,2}(z)$ in this more general situation.


\bibitem{mattsson}  W. Kohn and A. E. Mattsson, Phys. Rev. Lett. {\bf 81}, 3487
(1998).

\bibitem{abra}  M. Abramowitz and I. A. Stegun, {\it Handbook of
Mathematical Functions} (Dover, New York, 1964).

\bibitem{jung}  J. Jung, J. E. Alvarellos, E. Chacon, and P.
Garcia-Gonzalez, J. Phys.: Condens. Matter {\bf 19}, 266008 (2007).

\bibitem{note}  The first approximation $k_{F}^{m}\left| z-z^{\prime
}\right| \simeq k_{F}^{m}z$ can be justified for $z\rightarrow \infty ,$
even considering that the $z^{\prime }$ integral covers the range from $%
-\infty $ to $+\infty .$ This is due to the fact that the main contribution to
this integral comes from values of $z^{\prime }$ inside the slab, so the range
of 
$z^{\prime }$ could be effectively restricted to the region $-d\lesssim
z^{\prime }\lesssim 0.$

\bibitem{note.bb} Note that $\overline{z}^m =-d/2$. So $\beta$ is always
 negative as a sum of two negative terms.

\bibitem{nastos}  Fred Nastos, (Ph.D.
Thesis, Queen's University, Kingston, Ontario, Canada, 2000).

\bibitem{new} C. M. Horowitz, C. R. Proetto, and J. M. Pitarke (unpublished). 

\bibitem{pitarke}  J. M. Pitarke and A. G. Eguiluz, Phys. Rev. B {\bf 57}, 6329
(1998); {\bf 63}, 045116 (2001).

\bibitem{formu.1}  We have taken then: 
$E^{unif}_{K}(d)= e^2 (3 \pi^2 n)^{5/3} d/(20 \pi^2 a_0) $, 
$E^{unif}_{el}(d)=0$, and 
$E^{unif}_{x}(d)=-3e^2 (3 \pi^2)^{1/3} (n)^{4/3} d /(4 \pi a_0)$, where 
$n$ is the density of the uniform electron gas. 

\bibitem{note1} By $x$-only LDA we mean that in the KS equations the actual
exchange potential $V_{x}(z)$ is replaced at this point by the exchange
potential of a uniform electron gas with the local density, i.e., 
$V_{x}^{\text{LDA}}(z)=-\left[ 6n(z)/\pi \right] ^{1/3}.$


\bibitem{zhang1}  See Ref.~\onlinecite{zhang}. Note that as the work function
is defined as the difference between two energies, it is independent of the
choice for the zero ef energy.

\bibitem{schulte1}  F. K. Schulte, J. Phys. C: Solid State Phys. {\bf 7},
L370 (1974).

\bibitem{schulte}  F. K. Shulte, Surf. Sci. {\bf 55}, 427 (1976).

\end{references}
\end{document}